\documentclass[preprint,3p]{elsarticle}
\usepackage{amsmath,amssymb}
\usepackage{xspace}
\usepackage{longtable}

\usepackage{ulem}
\usepackage{color}
\usepackage[mathscr]{eucal}
\usepackage{pifont}
\usepackage{listings}

\newcommand\La{\mathscr{L}}
\newcommand\SARAH{{\tt SARAH}\xspace}
\newcommand{\SARAHv}[1]{\SARAH~\texttt{#1}\xspace}
\newcommand\MG{{\tt MadGraph}\xspace}
\newcommand{\MGv}[1]{\MG~{\texttt{#1}}\xspace}
\newcommand\FeynArts{{\tt FeynArts}\xspace}
\newcommand\FormCalc{{\tt FormCalc}\xspace}
\newcommand\CalcHep{{\tt CalcHep}\xspace}
\newcommand\CompHep{{\tt CompHep}\xspace}
\newcommand\WHIZARD{{\tt WHIZARD}\xspace}
\newcommand\OMEGA{{\tt O'Mega}\xspace}
\newcommand\SPheno{{\tt SPheno}\xspace}
\newcommand\MO{{\tt MicrOmegas}\xspace}
\newcommand\HB{{\tt HiggsBounds}\xspace}
\newcommand\SSP{{\tt SSP}\xspace}
\newcommand{\Fortran}{\texttt{Fortran}\xspace}

\title{\SARAH 3.2: Dirac Gauginos, UFO output, and more}

\author[fs]{Florian Staub} 
\address[fs]{Bethe Center for Theoretical Physics \& Physikalisches Institut der 
 Universit\"at Bonn, \\
Nu{\ss}allee 12, 53115 Bonn, Germany}
\ead[fs]{fnstaub@th.physik.uni-bonn.de}

\begin{document}

\begin{flushright}
Bonn-TH-2012-17
\end{flushright}

\begin{abstract}
\SARAH is a Mathematica package optimized for the fast, efficient and precise
 study of supersymmetric models beyond the MSSM: a new model can be defined in a
 short form and all vertices are derived. 
 This allows \SARAH to
 create model files for \FeynArts/\FormCalc, \CalcHep/\CompHep and
 \WHIZARD/\OMEGA. The newest version of \SARAH\ now
 provides the possibility to create model files in the UFO format which is
 supported\ by \MGv{5}, \texttt{MadAnalysis 5}, \texttt{GoSam}, and soon by
 \texttt{Herwig++}. Furthermore, \SARAH\ also calculates  the
 mass matrices, RGEs and 1-loop corrections to the mass spectrum. This
 information is used to write source code for \SPheno in order to create a
 precision spectrum generator for the given model. This
 spectrum-generator-generator functionality as well as the output of \WHIZARD
 and \CalcHep model files have seen further improvement
 in this version. Also models including Dirac Gauginos are supported with the new version of \SARAH,
 and additional checks for the consistency of the implementation of new models have been created.
\end{abstract}

\maketitle

\section*{Program Summary}
{\bf Manuscript Title}: SARAH 3.2: Dirac Gauginos, UFO output, and more \\
{\bf Author}: Florian Staub \\
{\bf Program title}: SARAH \\
{\bf Programming language}: Mathematica \\
{\bf Computers for which the program has been designed}: All Mathematica
is available for \\
{\bf Operating systems}: All Mathematica is available for \\
{\bf Keywords}: Supersymmetry, vertices, model files, Madgraph, UFO, model building \\
{\bf CPC Library Classification}: 11.1, 11.6 \\
{\bf Reasons for the new version}: SARAH provides with this version the possibility to create model files in the UFO format. The UFO format is supposed to become a standard format for model files which should be supported by many different tools in the future. Also models with Dirac gauginos weren't supported in earlier versions.  \\
{\bf Nature of problem}: To use Madgraph for new models it is necessary to provide the corresponding model files which include all information about the interactions of the model. However, the derivation of the vertices for a given model and putting those into model files which can be used with Madgraph is usually very time consuming. \\
Dirac gauginos are not present in the minimal supersymmetric standard model (MSSM) or many extensions of it. Dirac mass terms for vector superfields lead to new structures in the supersymmetric (SUSY) Lagrangian (bilinear mass term between gaugino and matter fermion as well as new D-terms) and modify also the SUSY renormalization group equations (RGEs). The Dirac character of gauginos can change the collider phenomenology. In addition, they come with an extended Higgs sector for which a precise calculation of the 1-loop masses hasn't happened so far. \\
{\bf Solution method}: SARAH calculates the complete Lagrangian for a given model whose gauge sector can be any direct product of SU(N) gauge groups. The chiral superfields can transform as any, irreducible representation with respect to these gauge groups and it is possible to handle an arbitrary number of symmetry breakings or particle rotations.  Also the gauge fixing are automatically added. Using this information, SARAH derives all vertices for a model. These vertices can be exported to model files in the UFO which is supported by Madgraph and other codes like GoSam, MadAnalysis or ALOHA. 
\\
The user can also study models with Dirac gauginos. In that case SARAH includes all possible terms in the Lagrangian stemming from the new structures and can also calculate the RGEs. The entire impact of these terms is then taken into account in the output of SARAH to UFO, CalcHep, WHIZARD, FeynArts and SPheno. \\
{\bf Unusual features}: SARAH doesn't need the Lagrangian of a model as input to calculate the vertices. The gauge structure, particle and content and superpotential as well as rotations stemming from gauge symmetry breaking are sufficient. All further information derives SARAH by its own. Therefore, the model files are very short and the implementation of new models is fast and easy. In addition, the implementation of a model can be checked for physical and formal consistency. In addition, SARAH can generate Fortran code for a full 1-loop analysis of the mass spectrum in the context for Dirac gauginos. \\
{\bf Restrictions}: SARAH can only derive the Lagrangian in automatized way for N=1 SUSY models, but not for those with more SUSY generators. Furthermore, SARAH supports only renormalizable operators in the output of model files in the UFO format and also for CalcHep, FeynArts and WHIZARD. Also color sextets are not yet included in the model files for Monte Carlo tools. Dimension 5 operators are only supported in the calculation of the RGEs and mass matrices.  \\
{\bf Summary of revisions}: Support of models with Dirac gauginos. Output of model files in the UFO format, speed improvement in the output of WHIZARD model files, CalcHep outputs supports the internal diagonalization of mass matrices, output of control files for LHPC spectrum plotter, support of generalized PDG numbering scheme PDG.IX, improvement of the calculation of the decay widths and branching ratios with SPheno, the calculation of new low energy observables are added to the SPheno output, the handling of gauge fixing terms has been significantly simplified.  \\
{\bf Does the new version supersede the previous version?}:  Yes, the new version includes all known features of previous version but provide also the new features mentioned above. \\
{\bf Running time}: Measured CPU time for the evaluation of the MSSM using a Lenovo Thinkpad X220 with i7 processor (2.53 GHz). Calculating the complete Lagrangian: 9 seconds. Calculating all vertices: 51 seconds. Output of the UFO model files: 49 seconds.

\section{Introduction}
Supersymmetry (SUSY) is still one of the best-motivated extensions of the
 standard model (SM) of particle physics. Even the
 minimal supersymmetric standard model (MSSM) predicts many new
 particles \cite{Nilles:1983ge,Martin:1997ns} and has been studied
 intensively for decades. However, so far no hint
 of any of these particles has been found at the LHC
 \cite{Aad:2011ib,ATLAS:2011ad,Aad:2011cwa,Chatrchyan:2011ek,Chatrchyan:2011qs,Chatrchyan:2011zy}. On the other hand, 
 a particle around 125~GeV with properties fitting to the standard model Higgs boson has been detected \cite{Atlas:2012gk,CMS:2012gu}. 
 Even
 if this mass can be obtained within the MSSM or the
 constrained MSSM (CMSSM), the interest in non-minimal models has
 grown. Models with an extended Higgs sector can not only enhance
 the mass of the SM-like Higgs boson
 (e.g.  \cite{Ma:2011ea,Ellwanger:2011aa}), but
 also lead to a rich collider phenomenology
 (e.g. \cite{Krauss:2012ku,FileviezPerez:2012mj}) and possibly to new dark matter candidates
 (e.g. \cite{Cerdeno:2009dv,Burell:2011wh}).  Therefore, it is very interesting to
 study the phenomenological aspects of many extensions of the MSSM. However, the 
 implementation of new models in computer tools to calculate  mass
 spectra, the dark matter relic
 density, or to perform collider studies is usually a tedious work. 

The Mathematica package \SARAH was created to close this gap between 
 model building and the usage of well-known computer tools to get
 precise numerical results for new models. The main idea of
 \SARAH is that a new supersymmetric model can be easily and quickly implemented
 by just defining the gauge structure, particle content and
 superpotential, along with the gauge symmetry
 breaking and the corresponding field rotations. \SARAH\ 
 uses this information to derive the mass matrices, tadpole
 equations, renormalization group equations (RGEs) and vertices.
 Expressions for the 1-loop corrections to the 1- and
 2-point functions are also calculated by \SARAH.
 Since version \texttt{1.0},
 \SARAH \cite{Staub:2008uz,Staub:2009bi,Staub:2010jh} has supported
  the output of the vertices to model files which can be used
 with \CalcHep/\CompHep \cite{Pukhov:2004ca,Boos:1994xb} as well as
 \FeynArts/\FormCalc \cite{Hahn:2000kx,Hahn:2009bf}.
 Support for
 \WHIZARD/\OMEGA \cite{Kilian:2007gr,Moretti:2001zz} was added
in version \texttt{3.0}. 

In addition, \SARAH\ has become the first available
`spectrum - generator - generator': all information derived by
 \SARAH necessary for a 2-loop RGE evaluation and the calculation of the 1-loop
 corrected mass spectrum can be exported to \Fortran
 code which can then be compiled with \SPheno
 \cite{Porod:2003um,Porod:2011nf} to generate a fully-fledged spectrum
 generator for a given model. This interface between \SPheno and \SARAH allows 
 to study many aspects of a new model: it enables to constrain and relate parameters 
 due to the embedding in a GUT model. In addition, it covers also other aspects 
 generally important for parameter studies like the check of precision observables and 
 a precise mass spectrum calculation including the entire 1-loop corrections to all masses  
 for the given model.
 The {\tt SUSY Toolbox} \cite{Staub:2011dp}
combines all the possible outputs of \SARAH to create
 an environment for the study of SUSY models beyond the MSSM,
 consisting of the powerful public tools \SPheno,
 \CalcHep, \MO \cite{Belanger:2006is}, \HB \cite{Bechtle:2008jh,Bechtle:2011sb},
 \WHIZARD and \SSP \cite{Staub:2011dp}. 

The newest version of \SARAH continues this development.  The output of model
 files in the  UFO (Universal FeynRules output) format
 \cite{Degrande:2011ua} is now supported. Thus far, it 
 has only been possible to create UFO model files in an
 automatized way by using {\tt FeynRules} \cite{Christensen:2008py,Christensen:2009jx,Duhr:2011se,Fuks:2012im}. 
 This
 format can be read for example by \MGv{5} \cite{Alwall:2011uj}, {\tt GoSam} \cite{Cullen:2011ac}, 
 {\tt MadAnalysis} \cite{Conte:2012fm} and {\tt Aloha} \cite{deAquino:2011ub}. Hence it is now
 possible to study all models already implemented by \SARAH also with \MG.
 Therefore, \MG\ has recently also become part of
 the {\tt SUSY Toolbox}. In addition, there are other improvements in
 \SARAHv{3.1}: the outputs for \SPheno, \CalcHep and \WHIZARD\ 
 have been amended, the PDG.IX scheme is supported and the handling
 of gauge fixing terms has been automatized. Finally, control files for the
 {\tt LHPC Spectrum Plotter} can be written. 
 
 In addition, with \SARAHv{3.2} also models with Dirac gauginos are full supported \cite{Fox:2002bu,Hall:1990hq}, and checks 
 for the consistency of the implementation of a models have been added. These new routines check physical aspects of the model, like gauge anomalies,
  as well as formal aspects of the implementation in \SARAH. 
 
 This paper is organized as follows: in sec.~\ref{sec:implementation} we give information about the derivation
 of the SUSY Lagrangian, the supported Lorentz and color structures and discuss the new support of Dirac gauginos. 
 In addition, we present the functionality to perform checks of the implementation of model. In sec.~\ref{sec:calculations} 
 we recapitulate the physical information which \SARAH can derive for a given model.
 Sec.~\ref{sec:output} is dedicated to the possible output of \SARAH. The focus will be one the new interface to
 write model files in the UFO format. In addition, also changes in the interface to \SPheno, \WHIZARD and 
 \CalcHep are presented. In sec.~\ref{sec:newfeatures} we discuss the other new features in version {\tt 3.1} and {\tt 3.2}. 
 We conclude in  sec.~\ref{sec:conclusion}.

\section{Implementation and  checks of models in \SARAH}
\label{sec:implementation}
\subsection{The SUSY Lagrangian with Dirac gauginos}
\subsubsection{Derivation of the SUSY Lagrangian}
\label{sec:Lagrangian}
In general, already the Lagrangian  for the gauge eigenstates of a SUSY model is very lengthy and consists of (i) gauge boson self-interactions, (ii) interactions between gauginos and gauge bosons, (iii) kinetic and gauge interaction terms for fermions and scalars, (iv) interactions between gauginos, fermions and sfermions, (v) D-terms, (vi) gauge-fixing terms, (vii) soft-breaking masses for chiral and vector superfields, (viii) fermion interactions stemming from the superpotential, (ix) F-terms, and (x) the soft-breaking counterparts to the superpotential interactions. One aim of \SARAH is to automatize as many steps as possible in the implementation of a new model. Therefore, \SARAH uses the the method explained in Ref.~\cite{Martin:1997ns} to generate all of these terms from a minimal input: (i) and (ii) are already completely fixed by the definition of the gauge group. To get (iii)--(vii) the user just has to define the chiral superfields as well as their quantum numbers with respect to the gauge groups. The transformation properties and the corresponding generators for non-fundamental, irreducible representations are derived by \SARAH using Young tableaux and don't demand any further input from the user than the dimensions or (optionally) the Dynkin labels. Also the impact of kinetic mixing is completely taken into account by including off-diagonal gauge couplings in the presence of several Abelian gauge groups \cite{Fonseca:2011vn}. (viii)--(x) are based on the superpotential which the user can add in a compact form as shown in \ref{app:model}. 

By default, \SARAH generates always the most general Lagrangian including all of these terms with the most general flavor structure. However, it might be that the user is interested, for instance, in a SUSY breaking scenario in which not all the standard soft terms arise. For this purpose distinct flags exist  to suppress specific structures:
\begin{itemize}
 \item \verb"AddTterms = True/False;", default: \verb"True", includes/excludes trilinear softbreaking couplings
 \item \verb"AddBterms = True/False;", default: \verb"True", includes/excludes bilinear softbreaking couplings
 \item \verb"AddLterms = True/False;", default: \verb"True", includes/excludes linear softbreaking couplings
 \item \verb"AddSoftScalarMasses = True/False;", default: \verb"True", includes/excludes soft-breaking scalar masses
 \item \verb"AddSoftGauginoMasses = True/False;", default: \verb"True", includes/excludes Majorana masses for gauginos
 \item \verb"AddDiracGauginos = True/False;", default: \verb"False", includes/excludes Dirac masses for gauginos
 \item \verb"AddSoftTerms = True/False;", default: \verb"True", includes/excludes all soft-breaking terms
 \end{itemize}
Furthermore, the quartic interactions of scalars are by default related to the gauge and Yukawa couplings via the D- and F-terms. However, these relations can be spoiled due to threshold effects \cite{Mahbubani:2004xq}. Also in the non-supersymmetric limit of a model, it might be preferable to treat the quartic interactions as free terms. In this case, the user can disable the F- and D-terms via 
\begin{itemize}
 \item \verb"AddDterms = True/False;", default: \verb"True", includes/excludes all D-terms
 \item \verb"AddFterms = True/False;", default: \verb"True", includes/excludes all F-terms
\end{itemize}
and define the quartic interactions separately. Using this method, it is possible to implement also  non-supersymmetric models in \SARAH. Two examples for such an implementation are already included in \SARAH: the SM and an inert Higgs doublet model. \\

In the last years another structure had been become popular in SUSY model building: Dirac mass terms for gauginos. These are bilinear terms between a vector superfield and a chiral superfield in the adjoint representation. Expanding these terms in component fields leads to two physical relevant terms in the Lagrangian  \cite{Benakli:2011vb}:
\begin{equation}
\label{eq:DG}
\La_{GF} = - m_D \lambda_a \psi_a + \sqrt{2} m_D \phi_a D_a 
\end{equation}
The first term is the Dirac mass term of the gaugino $\lambda_a$ and the fermion in the adjoint representation $\psi_a$, the second term are new D-term contributions involving the auxiliary part $D_a$ of the vector superfield and the scalar component $\phi_a$ of the chiral superfield. Models with Dirac gauginos are now fully supported in \SARAH: both terms of eq.~(\ref{eq:DG}) are included in the Lagrangian and the possible impact of kinetic mixing  is also taken into account in this context \cite{Benakli:2009mk}. Furthermore, the corresponding RGEs are calculated at 2-loop level, see sec.~\ref{sec:RGEs}. Dirac gauginos  are not included by default just because it is still more common to study models like the NMSSM without those terms. Hence, to include Dirac mass terms for gauginos  the user has to add the flag
\begin{verbatim}
AddDiracGauginos = True;
\end{verbatim}
to the model file. \SARAH uses \verb"MD<>v<>c" as name for the new mass parameters where \verb"v" is the name of the vector and \verb"c" the name of the chiral superfield, see \ref{app:model}. If several fields in the adjoint representation of one gauge group are present, \SARAH will generate the corresponding terms for all of them. To remove some of them, the parameters can be put to zero in the parameters file of the model definition, see also \ref{app:model}. %We give in \ref{app:model} an example for the implementation of a model with Dirac gauginos in \SARAH.  \\
% Note, \SARAH also calculate the RGEs for the new terms as well as the additional contributions to linear superpotential and soft-breaking, see  sec.~\ref{sec:RGEs}. 
 
\subsubsection{Supported Lorentz and color structures}
\label{sec:Lorentz}
The SUSY Lagrangian for a renormalizable model includes the  following Lorentz structures which are all fully supported by \SARAH: three and four scalar interactions ({\tt SSSS}, {\tt SSS}), interactions between two scalars and one or two vector bosons ({\tt SSV}, {\tt SSVV}), interactions between two vector bosons and one scalar ({\tt SVV}), interactions of ghost fields with scalars and vector bosons ({\tt GGS}, {\tt GGV}), interactions of fermions with scalar and vector bosons ({\tt FFS}, {\tt FFV}) as well as self-interactions between three and four vector bosons ({\tt VVV}, {\tt VVVV}). 

In addition, higher dimensional terms are partially supported by \SARAH: the user can add also terms with four superfields to the superpotential. \SARAH calculates also for these terms the 2-loop RGEs. Furthermore, it adds to the Lagrangian the dimension 5 operators stemming from these terms ({\tt FFSS} interactions), i.e. the potential impact to fermion masses is taken into account. However, the operators involving six scalars are not added to the Lagrangian. This is also the case for the operators stemming from those {\tt SSSSSS}-operators in the case that same scalar receive a vacuum expectation value.  In addition, higher dimensional operators are not yet included in the output of model files for Monte Carlo (MC) tools or \FeynArts. 

As written above, \SARAH uses Young tableaux to derive automatically the transformation properties of any irreducible representation of any $SU(N)$ gauge group. This information is also used to contract terms in the superpotential using Kronecker Deltas as well as the Levi-Civita Tensor. Hence, \SARAH treats non-fundamental representations  always as tensor products of the fundamental representation. It supports internally all possible color structures, for instance, in the calculation of the RGEs. However, in the output of model files for MC tools only the color singlet, triplet and octet are supported. In addition, only the UFO format can handle so far terms transforming as ${\bf3}\otimes{\bf3}\otimes{\bf3}$ under $SU(3)$ while this is not the case for \CalcHep or \WHIZARD. Since \CalcHep uses only implicit color indices, the color flow for 4-point vertices would suffer from an ambiguity. Therefore, \SARAH splits these terms automatically in two 3-point interactions introducing new auxiliary fields as this is usually done in \CalcHep model files.

\subsection{Checking models}
As already mentioned above, \SARAH tries to minimize the input of the user to a minimal amount. This also reduces the possibility of input errors. In addition new routines have been added to \SARAH to check the model itself as well as the model files for internal consistency. The following checks are performed when the function
\begin{verbatim}
 CheckModel
\end{verbatim}
is called:
\begin{itemize}
 \item Leads the particle content to gauge anomalies?
 \item Leads the particle content to Witten anomalies?
 \item Are all terms in the superpotential in agreement with charge, and if defined, $R$-parity conservation?
 \item Are there other, renormalizable terms allowed in the superpotential by gauge invariance beside those defined by the user. 
 \item Are all mixings between the fields consistent with unbroken gauge groups?
 \item Are all definitions of Dirac spinors consistent with unbroken gauge groups?
 \item Are there additional bilinear terms in the Lagrangian of the mass eigenstates which can cause additional mixing between fields?
 \item Are all mass matrices irreducible? (Otherwise a mixing between fields has been assumed which is not present)
 \item Are the properties of all particles and parameters defined correctly?
\end{itemize}
Note, the check for anomalies works so far only for supersymmetric models. 

\section{Calculations performed by \SARAH}
\label{sec:calculations}
We are going to present briefly the physical information which \SARAH can derive for a given model. For more information as well as detailed examples we refer the interested reader to Refs.~\cite{Staub:2008uz,Staub:2010jh} as well as the \SARAH manual. 
\subsection{Mass matrices and tadpole equations}
When the complete Lagrangian is calculated, tree level relations can easily be extracted: the mass matrices and tadpole equations are derived automatically for each set of eigenstates during the evaluation of a model. The user has access to both information by using the command
\begin{verbatim}
 MassMatrix[Particle]
\end{verbatim}
for the mass matrix of the mass eigenstate  \verb"Particle" and 
\begin{verbatim}
 TadpoleEquation[X]
\end{verbatim}
for the tadpole equation corresponding to a scalar eigenstate {\tt X}. 

For instance, {\tt Sd} is the name for the d-squarks and {\tt phid} is the CP even component of the neutral down-type Higgs,  see \ref{app:model}. Hence, {\tt MassMatrix[Sd]} returns the $6\times 6$ mass matrix for the d-squarks, while {\tt TadpoleEquation[phid]} returns the vacuum condition $\partial V/\partial \phi_d = 0$. \\
In the gauge sector it is common to give different names to the particles stemming from one mass matrix instead of adding a generation index. In this case it is sufficient to give one mass eigenstate as input: {\tt MassMatrix[VP]} as well as {\tt MassMatrix[VZ]} return the $2\times 2$ mass matrix for the neutral gauge bosons. 

\subsection{Vertices}
\label{sec:vertices}
The vertices can be calculated in two ways. Either it is possible to calculate the vertices for a specific choice of external particles or to calculate all vertices of the complete model at once. The former task is evolved by
\begin{verbatim}
 Vertex[{Particles},Options];
\end{verbatim}
The argument of this function is a list with the external particles. The options are
\begin{itemize}
\item \verb"Eigenstates", Value: \verb"$EIGENSTATES", Default: final eigenstates \\ 
Fixes the considered eigenstates, since these are not necessarily uniquely defined by the considered particles 
\item \verb"UseDependences": Value \verb"True" or \verb"False", Default: \verb"False" \\
Optional relations between the parameters which have been defined will be used, if \verb"UseDependences" is set to \verb"True".   
\end{itemize} 
In the results, the different parts of a vertex are ordered by their Lorentz structures.  If possible, the expressions are simplified by using the unitarity of rotation matrices, the properties of generators of $SU(N)$ gauge groups and, if defined, simplifying assumptions about parameters. \\

A few examples to use {\tt Vertex} are in order:
\begin{itemize}
 \item \verb"Vertex[{bar[Fd], Fd, VZ}]": returns the vertex between the down-quarks and the $Z$-boson.
 \item \verb"Vertex[{bar[Fd], Fd, VZ},UseDependences->True]": returns again the vertex between the down-quarks and the $Z$-boson, but with $g_1$ and $g_2$ replaced by the electric charge $e$. 
 \item \verb"Vertex[{fB, FdL,conj[SdL]},Eigenstates->GaugeES]": returns the vertex for the gauge eigenstates bino--left-down-quark--left-down-squark
\end{itemize}

All vertices for a set of eigenstates are calculated at once  by
\begin{verbatim}
MakeVertexList[$EIGENSTATES, Options];
\end{verbatim}
Here, first the eigenstates are defined and the possible options are
\begin{enumerate}
\item \verb"effectiveOperators", Values: \verb"True" or \verb"False", Default: \verb"False" \\
Defines, if also {\tt FFSS} vertices should be calculated.
\item \verb"GenericClasses", Values: \verb"All" or a list of generic types, Default: \verb"All"\\
Calculates the vertices only for the given types of interaction
\end{enumerate}
{\tt MakeVertexList} searches for all possible interactions present in the Lagrangian and creates lists for the generic subclasses of interactions, e.g. {\tt VertexList[FFS]} or {\tt VertexList[SSVV]} for all two-fermion-one-scalar interactions and all two-scalar-two-vector-boson interactions, respectively. 

For instance {\tt MakeVertexList[EWSB]} calculates all (renormalizable) interactions for the mass eigenstates after electroweak symmetry breaking (EWSB), while 
\begin{verbatim}
MakeVertexList[GaugeES,GenericClasses->{FFS,FFV}]
\end{verbatim}
would only calculate trilinear vertices involving fermions for the gauge eigenstates. \\

The calculation of the vertices has been cross-checked for several models with existing reference. For a discussion of these comparisons see Refs.~\cite{Staub:2009bi,Staub:2011dp} for the MSSM, Refs.~\cite{Liebler:2010bi,Ender:2011qh,Benbrik:2012rm} for the NMSSM, Ref.~\cite{Graf:2012hh} for the NMSSM with CP violation, and Ref.~\cite{Vormwald:ILC} for the  MSSM with bilinear $R$-parity violation. The $\mu\nu$SSM was cross checked by the authors of Ref.~\cite{Bartl:2009an}. For other models without the possibility of performing cross checks, the self consistency of the vertices has been checked by using the \SPheno interface discussed in sec.~\ref{sec:SPheno}: for instance, non-trivial checks are the relations between the Goldstone and vector boson masses at 1-loop.

\subsection{Renormalization group equations}
\label{sec:RGEs}
\SARAH calculates the SUSY RGEs at 1- and 2-loop level. This is done by using the generic formulas of Ref.~\cite{Martin:1993zk} which have been extended by the results of Ref.~\cite{Fonseca:2011vn} to include the entire impact of gauge kinetic mixing for models with several Abelian gauge groups. \\
Furthermore, also the changes in the RGEs in the presence of Dirac gaugino mass terms are included at the 2-loop level by using the results of Ref.~\cite{Goodsell:2012fm}. This includes the $\beta$-functions for the new mass parameters itself as well as new contribution to the RGEs of tadpole terms. The calculation of the RGEs can be started after the initialization of a model via
\begin{verbatim}
CalcRGEs[Options];
\end{verbatim}
The different options are
\begin{itemize}
\item \verb"TwoLoop", Value: \verb"True" or \verb"False", Default: \verb"True" \\
If also the two loop RGEs should be calculated.
\item \verb"ReadLists", Value: \verb"True" or \verb"False", Default: \verb"False" \\
If the RGEs have already be calculated, the results are saved in the output directory. The RGEs can be read from these files instead of doing the complete calculation again. 
\item \verb"VariableGenerations", Value: List of particles, Default: \verb"{}"\\
Some theories contain heavy superfields which should be integrated out above the SUSY scale. Therefore, it is possible to calculate the RGEs assuming the number of generations of specific superfields as free variable to make the dependence on these fields obvious. The new variable is named \verb"NumberGenertions[X]", where \verb"X" is the name of the superfield.
\item \verb"NoMatrixMultiplication", Values: \verb"True" or \verb"False", Default: \verb"False"\\
Normally, the \(\beta\)-functions are simplified by writing the sums over generation indices as matrix multiplication. This can be switched off using this option. 
\item \verb"IgnoreAt2Loop", Values: a list of parameters, Default: \verb"{}"\\
The calculation of 2-loop RGEs for models with many new interactions can be very
 time-consuming. However, often one is only interested in the dominant
 effects of the new contributions at the 1-loop level. Therefore, {\tt IgnoreAt2Loop -> \$LIST} can be used to neglect parameters at the 2-loop level. The entries of \$LIST can be superpotential or soft  SUSY-breaking parameters as well as gauge couplings.
\end{itemize}
The results of the calculation are saved as in previous versions in the arrays \verb"Gij" (Anomalous dimensions of all chiral superfields), \verb"BetaWijkl" (Quartic superpotential parameters), \verb"BetaYijk" (Trilinear superpotential parameters),  \verb"BetaMuij" (Bilinear superpotential parameters), \verb"BetaLi": (Linear superpotential parameters), \verb"BetaQijkl" (Quartic soft-breaking parameters), \verb"BetaTijk" (Trilinear soft-breaking parameters), \verb"BetaBij" (Bilinear soft-breaking parameters), \verb"BetaSLi" (Linear soft-breaking parameters),  \verb"Betam2ij" (Scalar squared masses), \verb"BetaMi" (Gaugino masses), \verb"BetaGauge" (Gauge couplings), and \verb"BetaVEVs" (VEVs). The newly calculated $\beta$-functions for the Dirac mass terms are stored in \verb"BetaDGi". \\

All entries of these arrays are three-dimensional: the first entry is the name of  the considered parameter, the second entry is the 1-loop \(\beta\)-function and the third one is the two loop \(\beta\)-function. For instance, the Yukawa couplings are saved in \verb"BetaYijk". The first entry consists of
\begin{verbatim}
BetaYijk[[1,1]]:  Ye[i1,i2] ,
\end{verbatim}
i.e. this entry contains the \(\beta\)-functions for the electron Yukawa coupling carrying two generation indices. The corresponding 1-loop \(\beta\)-function is stored in {\tt BetaYijk[[1,2]]} and reads
\begin{verbatim}
(-9*g1^2*Ye[i1,i2])/5-3*g2^2*Ye[i1,i2]+3*trace[Yd,Adj[Yd]]*Ye[i1,i2]+ 
  trace[Ye,Adj[Ye]]*Ye[i1, i2]+3*MatMul[Ye,Adj[Ye],Ye][i1, i2]
\end{verbatim}
As can be seen the expressions have been simplified by using traces ({\tt trace}) and matrix multiplication ({\tt MatMul}). However, if one wants to keep the sum over involved indices explicitly, he can use 
\begin{verbatim}
CalcRGEs[NoMatrixMultiplication->True];
\end{verbatim}
The 2-loop \(\beta\)-function is saved in \verb"BetaYijk[[1,3]]" but we skip it here because of its length. 

\subsection{1-loop tadpoles, self-energies and masses}
\label{sec:OneLoopSelf}
\SARAH calculates the analytical expressions for the 1-loop corrections to the tadpoles and the 1-loop self-energies for all particles. The executed steps are a generalization of the procedure applied in Ref.~\cite{Pierce:1996zz}: the calculations are performed in \(\overline{\mbox{DR}}\)-scheme using 't Hooft gauge and they are started for the different eigenstates through
\begin{verbatim}
CalcLoopCorrections[$EIGENSTATES];
\end{verbatim}
For instance, {\tt CalcLoopCorrections[EWSB]} calculates the loop corrections for the mass eigenstates after EWSB. The results are stored in {\tt Tadpoles1LoopSums[EWSB]} and {\tt SelfEnergy1LoopSum[EWSB]}.\\
{\tt Tadpoles1LoopSums[EWSB]} is a two-dimensional array, where the first column gives the name of the corresponding scalar, the second entry the 1-loop correction. Also {\tt SelfEnergy1LoopSum[EWSB]} is a two-dimensional entry. However, while the first column gives the mass eigenstates for which the self-energy is calculated, the content of the second column depends on the the type of the field: for scalars it contains the 1-loop self energy (\(\Pi(p^2)\)),  for fermions it contains another list with the 1-loop self energies for the different helicity (\(\Sigma^L(p^2)\),\(\Sigma^R(p^2)\), \(\Sigma^S(p^2)\)), and for vector bosons it contains the 1-loop, transversal self energy (\(\Pi^T(p^2)\)). For more details how this information can be used to calculate the 1-loop mass spectrum as well as about the conventions used for the loop integrals we refer to Refs.~\cite{Pierce:1996zz} and \cite{Staub:2010jh}. 

\section{Output of \SARAH}
\label{sec:output}
We give in the following an overview about the possible output of \SARAH using the derived information described in sec.~\ref{sec:calculations}. The focus will be on the new interface for UFO model files, but also the improvements in the interface to \WHIZARD, \CalcHep and \SPheno are shown. 
\subsection{Model files in the UFO format}
\subsubsection{The UFO format}
\label{sec:ufo}
\SARAH can write model files in the UFO format to implement new models in
 \MGv{5}. This format is also supported by other tools like {\tt GoSam}
 \cite{Cullen:2011ac}, {\tt Aloha} \cite{deAquino:2011ub} or \texttt{MadAnalysis 5} \cite{Conte:2012fm}, and soon by
 \texttt{Herwig++} \cite{Gieseke:2003hm,Gieseke:2006ga}. In general, the model files written by \SARAH include the
 entire flavor and CP structure of the model. For a model which is already
 implemented in \SARAH, it is rather easy to create the
 requested output by typing inside Mathematica
\begin{verbatim}
 << $SARAH/SARAH.m
Start["$MODEL"];
MakeUFO[Options];
\end{verbatim}
The meaning of these three steps is the following: first, \SARAH has to be
 loaded by using either the relative or absolute path \verb"$SARAH" to the
 \SARAH installation. The second step is to initialize the
 requested model, for instance the MSSM via
 {\tt \$MODEL} = {\tt MSSM}. A list of all models which are delivered with the
 public version of \SARAHv{3.1} is given in \ref{app:models} or
 can be printed by \SARAH using the command {\tt ShowModels}. For a
 step-by-step explanation how to implement new models, we refer to
 Ref.~\cite{Staub:2009bi} and \cite{Staub:2011dp}.\\
During the initialization of a model \SARAH calculates the entire Lagrangian
 density for the mass eigenstates as well as the mass matrices and
 tadpole equations. 
When this is finished, the routine to write the model files in the UFO format is
 called. \SARAH will always generate the UFO model files for the last eigenstates
 defined in the model file which are the mass eigenstates after EWSB for all
 models contained in the public version.
 
 As optional argument ({\tt Exclude -> \$LIST}) 
 the user can define a list of generic vertices which 
 shouldn't be exported to the model files: see sec.~\ref{sec:Lorentz} for a list of
 all possible Lorentz structures. This might be helpful to speed up the output
 of the model files and to the keep the model files short and faster. Of course, 
 this option has to be used very carefully and the user has to be sure that the excluded vertices
 are irrelevant for the desired studies. By default \$LIST = {\tt \{SSSS, GGS, GGV\}} is used. For instance, 
 to generate a UFO model file including only three-point interactions and no ghost terms, use
 \begin{verbatim}
 MakeUFO[Exclude->{SSSS,SSVV,VVVV,GGS,GGV}]; 
 \end{verbatim}
The output written by \SARAH consists of the files
\begin{itemize}
 \item {\tt particles.py}: contains the particles  present in the model
 \item {\tt parameters.py}: contains all parameters present in the model 
 \item {\tt lorentz.py}: defines the Lorentz structures needed for the vertices
\item {\tt vertices.py}: defines the vertices 
\item {\tt couplings.py}: expressions to calculate the couplings
\item {\tt coupling\_orders.py}: defines hierarchies for the coupling orders
\end{itemize}
These files are saved in the directory
\begin{verbatim}
 $SARAH/Output/$MODEL/$EIGENSTATES/UFO/
\end{verbatim}
For all models delivered with \SARAH\ {\tt \$EIGENSTATES} is just {\tt EWSB}. 
This directory contains also additional files which are model independent and
 can therefore be used with all models:  {\tt function\_library.py},
 {\tt object\_library.py}, {\tt \_\_init\_\_.py} and
 {\tt write\_param\_card.py}. These files were kindly provided by Olivier
 Mattelaer. \\
 
\subsubsection{Using the UFO model files of \SARAH with \MGv{5}} 
To use the model files with \MGv{5}, it is sufficient to copy all files
 to {\tt \$MADGRAPH/models/\$NAME}. Here, {\tt \$MADGRAPH} is the
 directory containing the local \MG installation and {\tt \$NAME} is a
 freely-chosen name of a new subdirectory. This
 directory name is used afterwards to load the model in \MG via 
\begin{verbatim}
 import model $NAME
\end{verbatim}
\MG has a list of pre-defined names for the particles of the SM and MSSM which
 are used by default. However, it could be that there are conflicts between
 these names and the names used by \SARAH in an extension of the MSSM. For
 instance, {\tt h3} is defined in \MG as the
pseudoscalar Higgs in the MSSM, but \SARAH uses it in
 the NMSSM for the third-heaviest scalar Higgs.
 In order to avoid such clashes, a model can be loaded
 using only the names defined by
 the UFO files via
\begin{verbatim}
 import model $NAME -modelname
\end{verbatim}

To use these model files for the calculation of cross sections, it is, of
 course, necessary to provide the numerical values of all masses and parameters.
 The necessary input can be obtained by using a \SPheno module created by \SARAH
 for  the given model, see also section~\ref{sec:SPheno}. The SLHA
 spectrum files written by this \SPheno module can directly be used with \MG. In
 this way, the chain \SARAH -- \SPheno -- \MG
 provides a direct link from model building to collider phenomenology.  
 %Of course, it is also
 %possible to use the model files of \SARAH with \MG without including \SPheno by providing all 
 %necessary input parameters at the SUSY scale. However, the advantage of 
 %using \SPheno even a SUSY scale input is that it can deliver the full 1-loop
 %mass spectrum and calculates precision observables, see sec.~\ref{sec:SPheno}.
 
\subsubsection{Validation}
As we have shown in sec.~\ref{sec:Lagrangian} \SARAH tries to reduce the user input to the minimal amount and calculated the entire Lagrangian from basic principle. In addition, as discussed in sec.~\ref{sec:vertices} the vertex calculation, hence also the derivation of the Lagrangian, have been cross-checked for several models, while for the other models the vertices have been checked for self-consistency. Therefore, the main task to create an interface between \SARAH and the UFO format is the translation of the information inside \SARAH to the conventions of the UFO format. In general, this procedure is the same for the MSSM as for all other models implemented in \SARAH because the MSSM already contains all supported Lorentz structures as well as possible difficulties like clashing arrows in the fermion flow. \\

Therefore, to validate the model files written by \SARAH in the UFO
 format, a list of $1 \to 2$ decays and $2\to 2$ processes
was calculated using \MGv{5.1.4.8.4}. The results
were compared with those calculated with the MSSM model files
 delivered with \MG. For all $1 \to 2$ decays there was always
 an exact numerical agreement. The results for the $2 \to 2$ scattering
 are shown in \ref{app:UFOvalidation}. It can be seen
 that all observed differences are smaller than the numerical error estimated by \MG. 
 
 In addition, we used the possibility of \MG to perform checks for a given process:
 \begin{itemize}
  \item \textit{permutation}: Checks that the model and \MGv{5} are working properly by generating permutations of the process and
   checking that the resulting matrix elements give the same value.
  \item \textit{gauge}: Checks that processes with massless gauge bosons are gauge invariant
  \item \textit{lorentz\_invariance}: Checks that the amplitude is lorentz invariant
 \end{itemize}
We run these checks for all processes listed in \ref{app:UFOvalidation} for the MSSM as well as for processes in the NMSSM. In addition, we checked a more involved model with an extended gauge sector, the B-L-SSM and the newly implement MRSSM with Dirac gauginos.

\subsection{\SPheno output}
\label{sec:SPheno}
\SARAH can use all derived information about a supersymmetric model to write
 \Fortran source code. This code can be compiled with \SPheno and modules for
 new models are created without the need to write any line of source code by
 hand. The features of the \SPheno modules written by \SARAH are a precise
 calculation of the mass spectrum using 2-loop RGEs, and 1-loop
 corrections to all masses. In addition, routines are written to calculate the
 decay widths and branching ratios for SUSY and Higgs particles. Also the
 calculation of a set of electroweak precision observables is included. All
 calculations are done with the most general CP and flavor structure.
 This functionality of \SARAH has been checked against existing spectrum
 calculators for the MSSM, see Ref.~\cite{Staub:2011dp}, and for the
 NMSSM a comparison with {\tt NMSSM-Tools} \cite{Ellwanger:2004xm,Ellwanger:2006rn} was done \cite{Staub:2010ty}.
 However, the \SPheno output works also for more complicated models
  like the minimal SUSY $B-L$ model \cite{O'Leary:2011yq}, 
 left-right symmetric models \cite{Esteves:2010si,Esteves:2011gk} 
 or models with inverse seesaw mechanism \cite{Hirsch:2012kv,Abada:2012cq}. 
To output the \SPheno code, the command
\begin{verbatim}
MakeSPheno[Options];
\end{verbatim}
has to be used after starting the model. The different options are
\begin{itemize}
 \item {\tt ReadLists -> True/False}: if the vertices and RGEs for the given
 model have already been calculated, these results can be used and a new
 calculation is skipped. The default value is {\tt False}.
 \item {\tt InputFile -> \$FILENAME}: basic properties of the \SPheno version
 such as boundary conditions at the GUT scale are defined in a
 separate file. The default name for that file is {\tt SPheno.m}. 
 \item {\tt TwoLoop -> True/False}: defines if the 2-loop RGEs should be
 calculated and included in the \Fortran code. The default value is {\tt True}.
 \item {\tt StandardCompiler -> \$COMPILER}: \SARAH writes a makefile to compile
 the generated code together with an existing \SPheno installation. The standard
 compiler defined in the makefile is usually {\tt gfortran} but can be changed
 by this flag. 
\end{itemize}
For more information in general about the \SPheno output of \SARAH, we
 refer the interested reader to Ref.~\cite{Staub:2011dp}. \\
The main improvements of the \SPheno output in version {\tt 3.2} are in the
 calculation of the low energy observables and of the decay widths.
\SPheno modules written by previous versions of \SARAH
 calculate lepton- and quark flavor violating observables: $b \to s \gamma$, $l_i \to  l_j \gamma$,
 $l_i \to 3 l_j$ (with $l = (e,\mu,\tau)$), the anomalous magnetic moment and electric dipole
 moments of the charged leptons, as well as 
  $\delta\rho=1-\rho=\frac{\Pi_{WW}(0)}{m_W^2}-\frac{\Pi_{ZZ}(0)}{m_Z^2}$ ($\Pi_{ZZ}$, $\Pi_{WW}$ are the self-energies of the massive vector bosons).
 This list has been extended now by
\begin{itemize}
 \item $\mu - e$ conversion in nuclei (Al, Ti, Sr, Sb, Au, Pb) based on the
 results of \cite{Arganda:2007jw}
 \item $\tau \to l P^0$ with a pseudoscalar meson $P^0$
 ($\pi^0$, $\eta$, $\eta'$) based on the results of \cite{Arganda:2008jj}
 \item $Z \to l_i \bar{l}_j$ calculated and implemented by Kilian Nickel  \cite{Dreiner:2012mx}
%  \item $B_{s,d} \to l_i \bar{l}_j$ calculated and implemented by Kilian Nickel  \cite{Dreiner:2012dh}
\end{itemize}
We want to pronounce that similar to the previously implemented calculation of 
$l_i \to 3 l_j$ also for  $\mu - e$ conversion and $\tau$ into meson decays not 
only the photonic contributions 
are included. Those are often assumed to be dominant, however, it has been shown
recently that especially the $Z$-penguin can be very important in extensions of the MSSM
and have to be taken into account \cite{Hirsch:2012ax,Dreiner:2012mx,Abada:2012cq},
 but
also the Higgs-penguins can give sizable contributions \cite{Abada:2011hm}. 
Therefore, in all calculations all contributions of 
$Z$- and Higgs-penguins as well as box diagrams are taken included. We have cross checked 
the calculation of these observables for the MSSM against the results of other public 
codes like {\tt superiso} \cite{superiso}, {\tt SUSY\_Flavor} \cite{susyflavor} and, of course, {\tt SPheno} itself. 
We found often good agreement, but processes are
implemented with different accuracy in different tools. For instance, the treatment
of NLO QCD corrections \cite{Bobeth:2001jm}, chiral resummation
\cite{Crivellin:2011jt}, or the considered diagrams don't agree necessarily. Therefore, 
a detailed comparison of flavor codes is beyond the scope of this paper but will happen elsewhere \cite{comparison}. \\

In addition, the calculation of all decays has been improved by performing an
 RGE evaluation of all couplings from the SUSY scale to the mass scale of the
 decaying particle. Previously, the values of the parameters at
 the SUSY scale were used in all decays. Furthermore, the
 calculation of the loop-induced decays of a Higgs particle into
 two photons and two gluons includes now also the dominant QCD
 corrections based on the results given in Ref.~\cite{Spira:1995rr}. This leads to the following
 precision precision in the calculations: for all  fermionic SUSY particle the two- and 
 three body decays are calculated at tree-level, while for squarks, sleptons and additional 
 heavy vector bosons the two-body decays are calculated. In the Higgs sector, possible decays 
 into two SUSY particles, leptons and massive gauge bosons are calculated at tree-level. 
 For two quarks in the final state the
 dominant QCD corrections due to gluons are included \cite{Spira:1995rr}. The loop induced decays into two photons
 and gluons are fully calculated at LO with the dominant NLO corrections as just mentioned. In 
 addition, in the Higgs decays also final states with off-shell gauge bosons ($Z Z^*$, $W W^*$) are 
 included.

\subsection{Output for \CalcHep and \WHIZARD}
\paragraph*{\CalcHep} \CalcHep is  able to read the numerical
 values of the masses and mixing matrices from a SLHA spectrum file or to take
 the values defined in the variable file; as an alternative, the
 \texttt{SLHA+} functionality \cite{Belanger:2010st} added routines to \CalcHep
 to diagonalize mass matrices to obtain the eigenvalues and eigenvectors by 
 itself. Therefore, we have implemented in \SARAH the option
 to include the mass matrices of a given value in the \CalcHep model files and
 to parametrize all vertices by the rotation matrices calculated
 internally by \CalcHep. To use this possibility, the
 \CalcHep output has to be started via
\begin{verbatim}
CHep[SLHAinput->False, CalculateMasses->True];
\end{verbatim}
The first option has to be used to disable the default approach that all masses
 and rotation matrices are taken from a SLHA spectrum file. The second option is
 used to write the necessary routines to the \CalcHep files to diagonalize all
 mass matrices. If this option were also set to
 {\tt False}, the masses and rotation matrices would be expected to
 be given in the variables file ({\tt varsX.mdl}) of \CalcHep. \\
Another, small improvement in the \CalcHep output is that \SARAH\ now
 adds  the electric charge of all particles in the {\tt Aux} column of
 the particle list. This is especially helpful for models with an extended gauge
 sector because \CalcHep might not be able to derive this information
on its own.

\paragraph*{\WHIZARD} The speed of the \WHIZARD output has 
 been improved significantly: to write the model files for the MSSM, the
 running time has been reduced by more than a factor of 2 
(from 12.2 to 5.0 minutes)\footnote{Time taken on a Lenovo Thinkpad X220 with 2.53 GHz}. 
For a more involved model like the $B-L$-SSM, the speed improvement is even larger (from 6.9 to 1.3 hours).

\section{Other new features since version 3.0}
\label{sec:newfeatures}
\subsection{Gauge fixing terms} 
\label{sec:gaugefixing}
If ghost vertices were to be calculated by
 previous versions of \SARAH, it was necessary
  to define the gauge fixing terms in $R_\xi$ gauge. However, the
 new version of \SARAH derives these terms automatically using the calculated
 kinetic terms of the Lagrangian. To this end, the condition is applied
 that the mixing between scalar particles and vector bosons vanishes.
 Afterwards, the derived gauge fixing terms are used to calculate the ghost
 interactions. \\
Since it can happen in models with an extended gauge sector that several
 Goldstone bosons are a mixture of the same gauge eigenstate, for each massive
 vector boson, the corresponding Goldstone boson has to be defined
 \begin{verbatim}
 {{   Description -> "Z-Boson",
          ...
          Goldstone -> Ah[{1}]}}, 
 ...
 {{   Description -> "Z'-Boson",
          ...
          Goldstone -> Ah[{2}]}}, 
 \end{verbatim}

\subsection{Generalized PDG numbering scheme}
Recently, a generalized PDG numbering scheme (PDG.IX) was proposed which might
 be helpful especially for extensions of the SM which include many
 new particles \cite{Brooijmans:2012yi,Basso:2012ew}. The basic idea is that the main properties
 of all particles (SM/Non-SM, spin, CP character, $B-L$ quantum number, electric
 charge, $SU(3)_C$ transformation) are encoded in a nine digit number.
 This provides not only information about the particles but fixes
 also the PDG particle code of new particles and 
 could help to prevent the use of randomly chosen
 numbers, and the confusion that this can cause. Therefore, it is now
 possible to define two different PDGs for a given particle in the model files
 in \SARAH:
\begin{verbatim}
 {{   Description -> "Gluino",
        ...
        PDG -> {1000021},
        PDG.IX ->{211110001}
        ...           }},
\end{verbatim}
By default, the entries of {\tt PDG} are used. To switch to the new scheme,
 either at the beginning of a \SARAH session or in the model files, the
 following statement has to be added:
\begin{verbatim}
UsePDGIX = True;
\end{verbatim}
In this case the new scheme will be used in the output for \SPheno, \WHIZARD,
\CalcHep and the UFO format. 

\subsection{Output for {\tt LHPC Spectrum Plotter }}
The LHPC spectrum plotter is a small but handy tool to produce plots of the SUSY mass
 spectrum based on the information given in a SLHA output file \cite{LHPC}. For examples of the 
 possible output see for instance Fig.~3 in Ref.~\cite{O'Leary:2011yq}. \\
For the output it is necessary to provide a second control file in addition to
 the SLHA spectrum file. The control file includes information about the paths
 to the necessary shell tools ({\tt gnuplot}, {\tt latex}, {\tt dvips},
 {\tt ps2eps}, {\tt rm}, {\tt mv}) and the \LaTeX{} name associated with a PDG
 number. In addition, the color and column used for the different particles are
 defined in that file. \SARAH can provide such a file which works nicely
 together with the spectrum file written by a \SPheno module also created by
 \SARAH. By default it assumes the standard paths under Linux, while the color
 and column of each particle can be defined in {\tt particles.m} using the new
 option {\tt LHPC}. For instance, to put the gluino in the fourth column and to
 use purple for the lines, the entry reads
\begin{verbatim}
 {{   Description -> "Gluino",
            ...
            LaTeX -> "\\tilde{g}",
            LHPC -> {4, "purple"},
            ... }},
\end{verbatim}
As name for the colors all available colors in {\tt gnuplot} can be used. The
 control file for a given set of eigenstates of the initialized model is written
 via
\begin{verbatim}
MakeLHPCstyle[$EIGENSTATES]; 
\end{verbatim}
and saved in the directory
\begin{verbatim}
$SARAH/Output/$MODEL/$EIGENSTATES/LHPC/ 
\end{verbatim}
It is used together with a spectrum file to create the figure by the shell
 command 
\begin{verbatim}
./LhpcSpectrumPlotter.exe SPheno.spc.$MODEL LHPC_$MODEL_Control.txt
\end{verbatim}

\section{Conclusion}
\label{sec:conclusion}
We have presented the improvements in the new version of the Mathematica package
 \SARAH. \SARAH was created to simplify the study of SUSY models beyond the MSSM
 and supported in earlier version the output of model files for
 \FeynArts/\FormCalc, \CalcHep/\CompHep, and \WHIZARD/\OMEGA as well as the
 source code output for \SPheno. In addition, the new version of \SARAH includes
 also model files in the UFO format which can be read by \MGv{5}. This
 significantly extends the list of SUSY models which can be studied with \MG at
 the moment. In this context, \MGv{5} has also been added to the
 {\tt SUSY Toolbox}. \\
 The models supported by \SARAH have been extended by the possibility to include
 Dirac mass terms for gauginos. In addition, routines to the implementation of models for 
 physical as well as formal consistency have been created.  
We have also shown that the handling of gauge fixing terms has been
 significantly simplified in \SARAH and the generalized PDG numbering scheme
 PDG.IX can now be used optionally. Furthermore, the speed of the \WHIZARD
 output has been improved and the \CalcHep output includes the possibility to
 diagonalize the mass matrices internally. The calculation of the decay widths
 and branching ratios in the \SPheno output has been improved and the list of
 calculated low energy constraints has been extended. Finally, \SARAH can write
 control files for the {\tt LHPC Spectrum Plotter}. 

\section*{Acknowledgments}
We thank Martin Hirsch, Ben O'Leary, Werner Porod, Laslo Reichert, Marco Pruna
 and  Avelino Vicente for their feedback and helpful suggestions during the
 development of \SARAH. For their support concerning the UFO format and \MG we
 thank Florian Bonnet, Olivier Mattelaer, Clausd Duhr, Benjamin Fuks and Johan
 Alwall. We thank Mark Goodsell and Karim Benakli for clarifying discussions about
 Dirac gauginos and cross checking the implementation of Dirac gauginos in \SARAH. Finally, 
 we thank Ben O'Leary also for his remarks on the first version of the manuscript. 

\begin{appendix}
\section{Model file for the MSSM/NMSSM with Dirac gauginos}
\label{app:model}
All information about a model are saved in three different files: \verb"Model.m", \verb"parameters.m" and \verb"particles.m". Only the first one is absolute necessary and contains the information about the gauge sector, particle content, superpotential and mixings. In \verb"parameters.m" properties to all parameters of the given model can be assigned (e.g. \LaTeX{} name, LesHouches entry, Real/complex). In \verb"particles.m" additional information about the particles are given, which might be needed for an appropriate output ($R$-parity, mass, width, PDG, \LaTeX{} name, output name). \\
We will concentrate here on a discussion of the model file itself and explain how Dirac gauginos are implemented. For more details about the implementation of a new model we refer the interested reader to Ref.~\cite{Staub:2010jh}. We will discuss the model presented in Refs.~\cite{Belanger:2009wf,Benakli:2010gi}. \\
\lstset{frame=shadowbox}
\begin{enumerate}
\item The gauge sector is \(U(1)\times SU(2)\times SU(3)\) and is defined
by declaring the corresponding vector superfields.
\begin{lstlisting}
Gauge[[1]]={B,   U[1], hypercharge, g1, False};
Gauge[[2]]={WB, SU[2], left,        g2, True};
Gauge[[3]]={G,  SU[3], color,       g3, False};
\end{lstlisting}
First, the name of the vector superfield is given. The second entry defines the
dimension of the group, the third one is the name of the gauge group and the
forth one the name of the corresponding gauge coupling. If the last entry is
set to {\tt True}, the sum over the charge induces is processed, otherwise the
charges are used as variable. In that case, the color charges are written as
indices, while the sum over isospins is expanded. \\
Note, \SARAH adds for every vector superfield automatically a soft-breaking
gaugino mass.
\item The next step is to define the matter sector. That's done by the array
{\tt Fields}. The conventions are the following. First, the root of the names
for the component fields is given (e.g. {\tt X}): the derived names of the
fermionic components start with {\tt F} in front (i.e. {\tt FX}), while for
scalars a {\tt S} is used (i.e. {\tt SX}). At second position the number of
generations is defined and the third entry is the name of the entire
superfield. The remaining entries are the transformation properties with
respect  to the different gauge groups. \\
Using these conventions, the doublet superfields \(\hat{q},\hat{l}, \hat{H}_d,
\hat{H}_u\) are added by
\begin{lstlisting}
Fields[[1]] = {{uL,  dL},  3, q,   1/6, 2, 3};  
Fields[[2]] = {{vL,  eL},  3, l,  -1/2, 2, 1};
Fields[[3]] = {{Hd0, Hdm}, 1, Hd, -1/2, 2, 1};
Fields[[4]] = {{Hup, Hu0}, 1, Hu,  1/2, 2, 1};
\end{lstlisting}
While for the singlet superfields \(\hat{d}, \hat{u}, \hat{e}\) 
\begin{lstlisting}
Fields[[5]] = {conj[dR], 3, d,  1/3, 1, -3};
Fields[[6]] = {conj[uR], 3, u, -2/3, 1, -3};
Fields[[7]] = {conj[eR], 3, e,    1, 1,  1};
\end{lstlisting}
is used. \\
The additional fields in the adjoint representation necessary to write down the Dirac mass terms are
\begin{lstlisting}
Fields[[8]] = {s, 1, S, 0, 1, 1};
Fields[[9]] = {{{T0/Sqrt[2],Tp},{Tm, -T0/Sqrt[2]}}, 1, T, 0, 3, 1};
Fields[[10]] = {Oc, 1, oc, 0, 1, 8}; 
\end{lstlisting}
As written in sec.~\ref{sec:Lagrangian} \SARAH treats non-fundamental representation as tensor products of the fundamental representation. 
Therefore, the entries of the  $SU(2)_L$ triplet superfield read
\begin{equation}
\hat{T} = \left(\begin{array}{cc} \hat{T}^0/\sqrt{2} & \hat{T}^+ \\ \hat{T}^- & - \hat{T}^0/\sqrt{2} \end{array} \right)
\end{equation}
Note, \SARAH adds also for scalars automatically the soft masses. 
\item The  command to include Dirac gaugino mass terms has to be added to the model file
\begin{lstlisting}
AddDiracGauginos = True;
\end{lstlisting}
This leads to the corresponding mass term and the D-terms. \SARAH names the new parameters using the corresponding superfield names as \verb"MDBS" (bino-singlet mass term), \verb"MDWBT" (wino-triplet mass term), \verb"MDGoc" (gluino-octet mass term). 
\item The superpotential of the model is given by
\begin{eqnarray}
W &=&  \hat{q} Y_u \hat{u} \hat{H}_u -  \hat{q} Y_d \hat{d} \hat{H}_d  - \hat{l}
   Y_e \hat{e} \hat{H}_d  +\mu \hat{H}_u \hat{H}_d + \lambda \hat{S} \hat{H}_d \hat{H}_u + \lambda_T \hat{H}_d \hat{T}  \hat{H}_u +\nonumber \\
\label{superpotential_MSSM}
 &&    L \hat{S} + \frac{1}{2} M_S \hat{S} \hat{S} + \frac{1}{3} \kappa \hat{S} \hat{S} \hat{S} + \frac{1}{2} M_T \hat{T} \hat{T} + \frac{1}{2} \lambda_S \hat{S} \hat{T} \hat{T} + \frac{1}{2} M_O \hat{O} \hat{O} 
\end{eqnarray}
and represented in \SARAH by
\begin{lstlisting}
SuperPotential = { {{1, Yu},{u,q,Hu}}, {{-1,Yd},{d,q,Hd}},
              {{-1,Ye},{e,l,Hd}}, {{1,\[Mu]},{Hu,Hd}},
              {{1,\[Lambda]},{S,Hd,Hu}}, {{1,LT},{Hd,T,Hu}},
              {{1,L1},{S}}, {{1/2,MS},{S,S}}, {{1/3, \[Kappa]},{S,S,S}}, 
              {{1/2,MT},{T,T}}, {{1/2,LS},{S,T,T}}, {{1/2,MO},{oc,oc}} 
  };
\end{lstlisting}
\item There are two different sets of eigenstates: the gauge eigenstates before
EWSB and the mass eigenstates after EWSB. The internal names are
\begin{lstlisting}
NameOfStates={GaugeES, EWSB};
\end{lstlisting}
\item The vector bosons and gauginos rotate after EWSB as follows
\begin{lstlisting}
DEFINITION[EWSB][GaugeSector]= 
{ {{VB,VWB[3]},{VP,VZ},ZZ},
  {{VWB[1],VWB[2]},{VWm,conj[VWm]},ZW},
  {{fWB[1],fWB[2],fWB[3]},{fWm,fWp,fW0},ZfW}
};    
\end{lstlisting}
The rotation matrices $Z^{\gamma Z}$ (\verb"ZZ"), $Z^W$ (\verb"ZW") and $Z^{\tilde{W}}$ (\verb"ZfW") are defined in the parameter file of the corresponding model as
\begin{equation}
Z^{\gamma Z} = \left(\begin{array}{cc} \cos\Theta_W & - \sin\Theta_W \\ \sin\Theta_W & \cos\Theta_W \end{array}\right)\,, \hspace{0.5cm} Z^W = \frac{1}{\sqrt{2}}\left(\begin{array}{cc} 1 & 1 \\ -i & i \end{array}\right)\,, \hspace{0.5cm} Z^{\tilde{W}} = \frac{1}{\sqrt{2}}\left(\begin{array}{ccc} 1 & 1 &0 \\ -i & i & 0 \\ 0 & 0 & 1 \end{array}\right) 
\end{equation}
This encodes the common mixing of vector bosons and gauginos after EWSB
\begin{align} 
W_{{1 \rho}} = & \,\frac{1}{\sqrt{2}} W^-_{{\rho}}  + \frac{1}{\sqrt{2}}
W^+_{{\rho}} \, , \hspace{1cm} 
W_{{2 \rho}} =  \,-i \frac{1}{\sqrt{2}} W^-_{{\rho}}  + i \frac{1}{\sqrt{2}}
W^+_{{\rho}} \\ 
W_{{3 \rho}} = & \,\cos\Theta_W  Z_{{\rho}}  + \sin\Theta_W  \gamma_{{\rho}}
\, , \hspace{1cm}  
B_{{\rho}} =  \,\cos\Theta_W  \gamma_{{\rho}}  - \sin\Theta_W  Z_{{\rho}} \\ 
\lambda_{{\tilde{W}},{1}} = & \,\frac{1}{\sqrt{2}} \tilde{W}^-  +
\frac{1}{\sqrt{2}} \tilde{W}^+ \, , \hspace{1cm}  
\lambda_{{\tilde{W}},{2}} =  \,-i \frac{1}{\sqrt{2}} \tilde{W}^-  + i
\frac{1}{\sqrt{2}} \tilde{W}^+ \, , \hspace{1cm}  
\lambda_{{\tilde{W}},{3}} =  \,\tilde{W}^0
\end{align} 

\item The neutral components of the scalar Higgs receive vacuum expectation
values (VEVs) \(v_d\)/\(v_u\) and split into scalar and pseudo scalar components. The same
happens for the singlet and the neutral component of the triplet
\begin{align} 
H_d^0 =  \, \frac{1}{\sqrt{2}} \left( \phi_{d}  + i \sigma_{d}  +  v_d  \right)
\, , \hspace{1cm} 
H_u^0 =  \, \frac{1}{\sqrt{2}} \left( \phi_{u}    + i  \sigma_{u}  +  v_u
\right)  \\
S =  \, \frac{1}{\sqrt{2}} \left( \phi_{S}  + i \sigma_{S}  +  v_S  \right)
\, , \hspace{1cm} 
T^0 =  \, \frac{1}{\sqrt{2}} \left( \phi_{T}    + i  \sigma_{T}  +  v_T
\right)
\end{align} 
This is encoded in \SARAH by
\begin{lstlisting}
DEFINITION[EWSB][VEVs]= 
{{SHd0,{vd,1/Sqrt[2]},{sigmad,I/Sqrt[2]},{phid,1/Sqrt[2]}},
 {SHu0,{vu,1/Sqrt[2]},{sigmau,I/Sqrt[2]},{phiu,1/Sqrt[2]}},
 {ST0, {vT, 1/Sqrt[2]}, {sigmaT, I/Sqrt[2]},{phiT,1/Sqrt[2]}},
 {Ss, {vS, 1/Sqrt[2]}, {sigmaS, I/Sqrt[2]},{phiS,1/Sqrt[2]}}};
\end{lstlisting}
\item The particles mix after EWSB to new mass eigenstates
\begin{lstlisting}
DEFINITION[EWSB][MatterSector]= 
{{{SdL, SdR           }, {Sd, ZD}},
 {{SuL, SuR           }, {Su, ZU}},
 {{SeL, SeR           }, {Se, ZE}},
 {{SvL                }, {Sv, ZV}},
 {{phid, phiu, phiS, phiT             }, {hh, ZH}},
 {{sigmad, sigmau, sigmaS, sigmaT     }, {Ah, ZA}},
 {{SHdm,conj[SHup],STm,conj[STp]      },   {Hpm,ZP}},
 {{fB, fW0, FHd0, FHu0, Fs, FT0       }, {L0, ZN}}, 
 {{{fWm, FHdm,FTm}, {fWp, FHup,FTp}}, {{Lm,UM}, {Lp,UP}}},
 {{{FeL},       {conj[FeR]}}, {{FEL,ZEL},{FER,ZER}}},
 {{{FdL},       {conj[FdR]}}, {{FDL,ZDL},{FDR,ZDR}}},
 {{{FuL},       {conj[FuR]}}, {{FUL,ZUL},{FUR,ZUR}}},
 {{fG,FOc},{GW,ZG}}}; 
\end{lstlisting}
This defines the mixings to the mass eigenstates: first, a list with gauge
eigenstates is given, then the name of the new mass eigenstates and the
mixing matrix follows. Hence, the first line is interpreted as
\begin{align} 
\tilde{d}_{L,{i \alpha}} = \sum_{j}Z^{D,*}_{j i}\tilde{d}_{{j \alpha}}\,, \hspace{1cm} 
\tilde{d}_{R,{i \alpha}} = \sum_{j+3}Z^{D,*}_{j i}\tilde{d}_{{j \alpha}}
\end{align} 
while the 9th line defines the mixing in the chargino sector
\begin{align} 
& \tilde{W}^- = \sum_j U^*_{j 1}\lambda^-_{{j}}\,, \hspace{1cm} 
\tilde{H}_d^- = \sum_j U^*_{j 2}\lambda^-_{{j}} \,, \hspace{1cm}
\tilde{T}_- = \sum_j U^*_{j 3}\lambda^-_{{j}} \,, \\
& \tilde{W}^+ = \sum_j V^*_{1 j}\lambda^+_{{j}}\,, \hspace{1cm} 
\tilde{H}_u^+ = \sum_j V^*_{2 j}\lambda^+_{{j}}, \hspace{1cm}
\tilde{T}^+ = \sum_j V^*_{3 j}\lambda^+_{{j}}
\end{align} 
In comparison to the MSSM the mass eigenstates change as follows: (i) there are 4 CP even Higgs states (defined by the 5th line of the above rotations); 
(ii) 4 CP odd Higgs states (6th line), from which the massless state is the Goldstone mode of the $Z$-boson;  (iii) 4 charged Higgs particles (7th line), from which the massless state is the Goldstone mode of the $W$-boson; (iv) six neutralinos (defined by the 7th); (v) three chargino as discussed above; (vi) two fermionic color octets due to the mixing of the gluino and 'octino' because of the Dirac mass term term (the Weyl spinors are denoted by $\tilde{g}_W$ (= \verb"GW")); (vii) one scalar color octet which doesn't mixing with any other particle. \\
Note, \SARAH uses the conventions of Ref.~\cite{Martin:1997ns} for the gaugino phase. Therefore, the definition
of the basis of neutralinos and charginos differs by a factor of $-i$ in comparison to Ref.~\cite{Skands:2003cj}. 
\item The Dirac spinors for the mass eigenstates are 
\begin{lstlisting}
DEFINITION[EWSB][DiracSpinors]={
 Fd - > {FDL, conj[FDR]},
 Fe  -> {FEL, conj[FER]},
 Fu  -> {FUL, conj[FUR]},
 Fv  -> {FvL, 0},
 Chi -> {L0, conj[L0]},
 Cha -> {Lm, conj[Lp]},
 Glu -> {GW, conj[GW]}
};
\end{lstlisting}
That leads to the replacements
\begin{equation}
d \rightarrow \left(\begin{array}{c} d_L \\ d_R \end{array} \right)\,, \dots \,,
\tilde{\chi}^- \rightarrow \left(\begin{array}{c} \lambda^- \\
(\lambda^+)^* \end{array} \right)\,, 
\tilde{g} \rightarrow \left(\begin{array}{c} \tilde{g}_W \\ \tilde{g}_W^*
\end{array} \right)\
\end{equation}
when going from four- to two-component formalism, or vice versa.
\end{enumerate}

\section{Models}
\label{app:models}
In \SARAHv{3.2} the following models are included:
\begin{itemize}
 \item Minimal supersymmetric standard model (see Ref.~\cite{Martin:1997ns} and references therein):  
    \begin{itemize} 
     \item With general flavor and CP structure ({\tt MSSM})
     \item Without flavor violation ({\tt MSSM/NoFV})
     \item With explicit CP violation in the Higgs sector ({\tt MSSM/CPV})
     \item In SCKM basis ({\tt MSSM/CKM})
    \end{itemize}
   \item Singlet extensions: 
   \begin{itemize}
    \item Next-to-minimal supersymmetric standard model ({\tt NMSSM},
 {\tt NMSSM/NoFV}, {\tt NMSSM/CPV}, {\tt NMSSM/CKM}) (see Refs.~\cite{Maniatis:2009re,Ellwanger:2009dp} and references therein)
    \item near-to-minimal supersymmetric standard model
 ({\tt near-MSSM}) \cite{Barger:2006dh}
    \item General singlet extended, supersymmetric standard model ({\tt SMSSM})  \cite{Barger:2006dh,Ross:2012nr}
  \end{itemize}
  \item Triplet extensions 
  \begin{itemize} 
    \item Triplet extended MSSM ({\tt TMSSM}) \cite{DiChiara:2008rg}
    \item Triplet extended NMSSM ({\tt TNMSSM}) \cite{Agashe:2011ia}
  \end{itemize}
   \item Models with $R$-parity violation  \cite{Hall:1983id,Dreiner:1997uz,Allanach:2003eb,Bhattacharyya:1997vv,Barger:1989rk,Allanach:1999ic,Hirsch:2000ef,Barbier:2004ez}
  \begin{itemize}
    \item bilinear RpV ({\tt MSSM-RpV/Bi}) 
    \item Lepton number violation ({\tt MSSM-RpV/LnV})
    \item Only trilinear lepton number violation ({\tt MSSM-RpV/TriLnV})
    \item Baryon number violation ({\tt MSSM-RpV/BnV})  
    \item $\mu\nu$SSM ({\tt munuSSM}) \cite{LopezFogliani:2005yw,Bartl:2009an}
  \end{itemize}
   \item Additional $U(1)'s$ 
  \begin{itemize}
    \item $U(1)$-extended MSSM ({\tt UMSSM})  \cite{Barger:2006dh}
    \item secluded MSSM ({\tt secluded-MSSM}) \cite{Chiang:2009fs}
    \item minimal $B-L$ model ({\tt B-L-SSM})  \cite{Khalil:2007dr,FileviezPerez:2010ek,O'Leary:2011yq,Basso:2012gz}
    \item minimal singlet-extended $B-L$ model ({\tt N-B-L-SSM})
  \end{itemize}
   \item SUSY-scale seesaw extensions
    \begin{itemize}
      \item inverse seesaw ({\tt inverse-Seesaw}) \cite{Malinsky:2005bi,Abada:2012cq,Dev:2012ru}
      \item linear seesaw ({\tt LinSeesaw}) \cite{Malinsky:2005bi,DeRomeri:2012qd}
      \item singlet extended inverse seesaw ({\tt inverse-Seesaw-NMSSM}) \cite{Gogoladze:2012jp}
      \item inverse seesaw with $B-L$ gauge group ({\tt B-L-SSM-IS})  \cite{Basso:2012ew}
      \item minimal $U(1)_R \times U(1)_{B-L}$ model with inverse seesaw
 ({\tt BLRinvSeesaw}) \cite{Hirsch:2011hg,Hirsch:2012kv}
\end{itemize}
 \item Models with Dirac Gauginos
   \begin{itemize}
    \item MSSM/NMSSM with Dirac Gauginos ({\tt DiracGauginos}) \cite{Belanger:2009wf,Benakli:2010gi,Benakli:2012}
    \item minimal $R$-Symmetric SSM ({\tt MRSSM}) \cite{Kribs:2007ac}
   \end{itemize}
 \item High-scale extensions
\begin{itemize}
 \item Seesaw 1 - 3 ($SU(5)$ version) ,
 ({\tt Seesaw1},{\tt Seesaw2},{\tt Seesaw3}) \cite{Borzumati:2009hu,Rossi:2002zb,Hirsch:2008dy,Esteves:2009vg,Esteves:2010ff}
 \item Left/right model ($\Omega$LR) ({\tt Omega}) \cite{Esteves:2010si,Esteves:2011gk}
\end{itemize}
\item Non-SUSY models:
\begin{itemize}
\item SM ({\tt SM}, {\tt SM/CKM}) (see for instance  Ref.~\cite{Hollik:2010id} and references therein)
\item inert Higgs doublet model ({\tt Inert}) \cite{LopezHonorez:2006gr}
\end{itemize}
\end{itemize}

\section{Validation of UFO output}
\label{app:UFOvalidation}
Cross section and numerical error calculated with \MGv{5} using the model files
 for the MSSM included in \MGv{5.1.4.8.4} ($\sigma_M$, $\delta_M$) and UFO model
 files written by \SARAH ($\sigma_S$, $\delta_S$). Here, $\sigma_i$ gives the value of
  the error estimation in  the cross section of \MG. $\delta$ is the relative uncertainty
  in percent calculated by $\delta_{S,M} = \frac{\Delta \sigma_{S,M}}{\sigma_{S,M}}$, where 
  $\Delta \sigma_{S,M}$ is the absolute
  error given by \MG. The difference was 
 calculated as $D = (\sigma_S - \sigma_M)/\sigma_S$.
 The run and parameter cards 
%   given in \ref{app:parcard} and \ref{app:runcard} have been used.
  given in \ref{app:cards} have been used.

\begin{longtable}{|l|cc|cc|c|} 
\hline 
Process & \(\sigma_S\) [fb] & \(\delta_S\)~[\%]&  \(\sigma_M\) [fb] & \(\delta_M\)~[\%] &  $D$~[\%]  \\ 
\hline 
\hline 
\(e \bar{e} \rightarrow W^- W^+\)                                                   & \(3.161\times 10^{-2}\) & \(2.24\times10^{-1}\) &  \(3.162\times 10^{-2}\) & \(2.63\times10^{-1}\) &  \(-2.66\times10^{-2}\) \\ 
\(\tau \bar{\tau} \rightarrow A^0 h\)                                               & \(1.783\times 10^{-7}\) & \(1.19\times10^{-1}\) &  \(1.783\times 10^{-7}\) & \(1.1\times10^{-1}\)  &  \(2.3\times10^{-2}\) \\ 
\(d \bar{d} \rightarrow \tilde{g} \tilde{g}\)                                       & \(6.229\times 10^{-2}\) & \(7.46\times10^{-2}\) &  \(6.23\times 10^{-2}\)  & \(7.4\times10^{-2}\)  &  \(-3.21\times10^{-3}\) \\ 
\(d \bar{d} \rightarrow \tilde{u}_2 \tilde{u}_2^*\)                                 & \(6.384\times 10^{-3}\) & \(7.52\times10^{-2}\) &  \(6.384\times 10^{-3}\) & \(7.52\times10^{-2}\) &  \(0.\) \\ 
\(e \bar{e} \rightarrow \tilde{u}_3 \tilde{u}_6^*\)                                 & \(3.94\times 10^{-5}\)  & \(6.65\times10^{-2}\) &  \(3.941\times 10^{-5}\) & \(6.67\times10^{-2}\) &  \(-2.79\times10^{-2}\) \\ 
\(e \bar{e} \rightarrow A^0 h\)                                                     & \(8.365\times 10^{-9}\) & \(6.67\times10^{-2}\) &  \(8.369\times 10^{-9}\) & \(7.54\times10^{-2}\) &  \(-5.02\times10^{-2}\) \\ 
\(\tilde{\chi}^0_1 \tilde{\chi}^0_1 \rightarrow W^- W^+\)                           & \(3.244\times 10^{-4}\) & \(2.02\times10^{-1}\) &  \(3.244\times 10^{-4}\) & \(2.02\times10^{-1}\) &  \(3.08\times 10^{-8}\) \\ 
\(d \bar{d} \rightarrow A^0 h\)                                                     & \(4.093\times 10^{-9}\) & \(6.67\times10^{-2}\) &  \(4.095\times 10^{-9}\) & \(7.54\times10^{-2}\) &  \(-5.13\times10^{-2}\) \\ 
\(\tilde{\chi}^0_1 \tilde{\chi}^0_1 \rightarrow h A^0\)                             & \(8.047\times 10^{-5}\) & \(6.43\times10^{-2}\) &  \(8.047\times 10^{-5}\) & \(6.43\times10^{-2}\) &  \(1.24\times 10^{-10}\) \\ 
\(d \bar{d} \rightarrow \tilde{\chi}^0_1 \tilde{\chi}^0_1\)                         & \(2.701\times 10^{-6}\) & \(8.54\times10^{-2}\) &  \(2.702\times 10^{-6}\) & \(7.61\times10^{-2}\) &  \(-3.29\times10^{-3}\) \\ 
\(\tilde{\chi}^0_1 \tilde{\chi}^0_1 \rightarrow Z Z\)                               & \(1.438\times 10^{-4}\) & \(2.54\times10^{-1}\) &  \(1.438\times 10^{-4}\) & \(2.54\times10^{-1}\) &  \(6.95\times 10^{-8}\) \\ 
\(\tilde{\chi}^0_1 \tilde{\chi}^0_1 \rightarrow \tilde{\chi}^0_1 \tilde{\chi}^0_1\) & \(9.924\times 10^{-6}\) & \(1.17\times10^{-1}\) &  \(9.924\times 10^{-6}\) & \(1.17\times10^{-1}\) &  \(0.\) \\ 
\(\tilde{\chi}^0_1 \tilde{\chi}^0_1 \rightarrow \tilde{\chi}^0_2 \tilde{\chi}^0_2\) & \(7.992\times 10^{-5}\) & \(1.58\times10^{-1}\) &  \(7.992\times 10^{-5}\) & \(1.58\times10^{-1}\) &  \(0.\) \\ 
\(\tilde{\chi}^0_1 \tilde{\chi}^0_1 \rightarrow \tilde{\chi}^0_3 \tilde{\chi}^0_3\) & \(2.653\times 10^{-2}\) & \(1.36\times10^{-1}\) &  \(2.653\times 10^{-2}\) & \(1.36\times10^{-1}\) &  \(0.\) \\ 
\(\tilde{\chi}^0_1 \tilde{\chi}^0_1 \rightarrow \tilde{\chi}^0_4 \tilde{\chi}^0_4\) & \(8.663\times 10^{-4}\) & \(1.35\times10^{-1}\) &  \(8.663\times 10^{-4}\) & \(1.35\times10^{-1}\) &  \(0.\) \\ 
\(\tilde{\chi}^0_1 \tilde{\chi}^0_1 \rightarrow H A^0\)                             & \(4.405\times 10^{-4}\) & \(4.94\times10^{-2}\) &  \(4.405\times 10^{-4}\) & \(4.94\times10^{-2}\) &  \(0.\) \\ 
\(\tilde{\chi}^0_1 \tilde{\chi}^0_1 \rightarrow \tilde{\chi}^+_1 \tilde{\chi}^-_1\) & \(5.918\times 10^{-2}\) & \(1.47\times10^{-1}\) &  \(5.918\times 10^{-2}\) & \(1.47\times10^{-1}\) &  \(-1.69\times10^{-3}\) \\ 
\(\tilde{\chi}^0_1 \tilde{\chi}^0_1 \rightarrow \tilde{\chi}^+_2 \tilde{\chi}^-_2\) & \(1.557\times 10^{-2}\) & \(1.62\times10^{-1}\) &  \(1.557\times 10^{-2}\) & \(1.62\times10^{-1}\) &  \(-6.42\times10^{-3}\) \\ 
\(\tilde{\chi}^0_1 \tilde{\chi}^0_1 \rightarrow \tilde{\chi}^0_1 \tilde{\chi}^0_2\) & \(5.641\times 10^{-5}\) & \(1.1\times10^{-1}\)  &  \(5.641\times 10^{-5}\) & \(1.1\times10^{-1}\)  &  \(0.\) \\ 
\(d \bar{d} \rightarrow \tilde{e}_1 \tilde{e}_1^*\)                                 & \(2.057\times 10^{-5}\) & \(6.77\times10^{-2}\) &  \(2.058\times 10^{-5}\) & \(8.71\times10^{-2}\) &  \(-4.67\times10^{-2}\) \\
\(e \bar{e} \rightarrow \tilde{e}_1 \tilde{e}_1^*\)                                 & \(4.46\times 10^{-3}\)  & \(4.71\times10^{-2}\) &  \(4.462\times 10^{-3}\) & \(3.69\times10^{-2}\) &  \(-5.91\times10^{-2}\) \\
\(e \bar{e} \rightarrow \tilde{e}_2 \tilde{e}_2^*\)                                 & \(1.509\times 10^{-4}\) & \(8.07\times10^{-2}\) &  \(1.51\times 10^{-4}\)  & \(8.67\times10^{-2}\) &  \(-8.42\times10^{-2}\) \\
\(e \bar{e} \rightarrow \tilde{e}_3 \tilde{e}_3^*\)                                 & \(1.319\times 10^{-4}\) & \(7.\times10^{-2}\)   &  \(1.32\times 10^{-4}\)  & \(8.6\times10^{-2}\)  &  \(-4.37\times10^{-2}\) \\ 
\(e \bar{e} \rightarrow \tilde{e}_5 \tilde{e}_5^*\)                                 & \(1.353\times 10^{-4}\) & \(7.8\times10^{-2}\)  &  \(1.354\times 10^{-4}\) & \(8.37\times10^{-2}\) &  \(-8.8\times10^{-2}\) \\ 
\(e \bar{e} \rightarrow \tilde{e}_6 \tilde{e}_6\)                                   & \(1.451\times 10^{-4}\) & \(7.75\times10^{-2}\) &  \(1.452\times 10^{-4}\) & \(8.19\times10^{-2}\) &  \(-8.14\times10^{-2}\) \\
\(e \bar{e} \rightarrow \tilde{e}_3 \tilde{e}_6\)                                   & \(4.568\times 10^{-6}\) & \(6.61\times10^{-2}\) &  \(4.569\times 10^{-6}\) & \(6.67\times10^{-2}\) &  \(-3.72\times10^{-2}\) \\
\(e \bar{e} \rightarrow \tilde{e}_1 \tilde{e}_4^*\)                                 & \(1.664\times 10^{-4}\) & \(1.02\times10^{-1}\) &  \(1.664\times 10^{-4}\) & \(9.05\times10^{-2}\) &  \(-6.18\times10^{-3}\) \\
\(e \bar{e} \rightarrow \tilde{u}_1 \tilde{u}_1^*\)                                 & \(2.715\times 10^{-4}\) & \(6.84\times10^{-2}\) &  \(2.718\times 10^{-4}\) & \(9.88\times10^{-2}\) &  \(-1.11\times10^{-1}\) \\
\(e \bar{e} \rightarrow \tilde{u}_2 \tilde{u}_2^*\)                                 & \(2.715\times 10^{-4}\) & \(6.84\times10^{-2}\) &  \(2.718\times 10^{-4}\) & \(9.88\times10^{-2}\) &  \(-1.11\times10^{-1}\) \\
\(e \bar{e} \rightarrow \tilde{u}_3 \tilde{u}_3^*\)                                 & \(1.686\times 10^{-4}\) & \(7.34\times10^{-2}\) &  \(1.688\times 10^{-4}\) & \(7.51\times10^{-2}\) &  \(-8.3\times10^{-2}\) \\ 
\(e \bar{e} \rightarrow \tilde{u}_4 \tilde{u}_4^*\)                                 & \(1.788\times 10^{-4}\) & \(7.65\times10^{-2}\) &  \(1.789\times 10^{-4}\) & \(7.97\times10^{-2}\) &  \(-7.44\times10^{-2}\) \\
\(e \bar{e} \rightarrow \tilde{u}_5 \tilde{u}_5^*\)                                 & \(1.788\times 10^{-4}\) & \(7.65\times10^{-2}\) &  \(1.789\times 10^{-4}\) & \(7.97\times10^{-2}\) &  \(-7.44\times10^{-2}\) \\
\(e \bar{e} \rightarrow \tilde{u}_6 \tilde{u}_6^*\)                                 & \(2.035\times 10^{-4}\) & \(7.1\times10^{-2}\)  &  \(2.037\times 10^{-4}\) & \(7.2\times10^{-2}\)  &  \(-9.73\times10^{-2}\) \\ 
\(e \bar{e} \rightarrow \tilde{u}_6 \tilde{u}_3^*\)                                 & \(3.94\times 10^{-5}\)  & \(6.65\times10^{-2}\) &  \(3.941\times 10^{-5}\) & \(6.67\times10^{-2}\) &  \(-2.79\times10^{-2}\) \\
\(e \bar{e} \rightarrow \tilde{\chi}^0_1 \tilde{\chi}^0_1\)                         & \(6.32\times 10^{-4}\)  & \(9.29\times10^{-2}\) &  \(6.316\times 10^{-4}\) & \(8.59\times10^{-2}\) &  \(6.87\times10^{-2}\) \\ 
\(e \bar{e} \rightarrow \tilde{\chi}^0_1 \tilde{\chi}^0_4\)                         & \(3.085\times 10^{-5}\) & \(9.78\times10^{-2}\) &  \(3.086\times 10^{-5}\) & \(9.03\times10^{-2}\) &  \(-4.42\times10^{-2}\) \\
\(e \bar{e} \rightarrow \tilde{\chi}^0_2 \tilde{\chi}^0_2\)                         & \(4.205\times 10^{-4}\) & \(7.35\times10^{-2}\) &  \(4.206\times 10^{-4}\) & \(7.36\times10^{-2}\) &  \(-3.36\times10^{-2}\) \\
\(e \bar{e} \rightarrow \tilde{\chi}^0_3 \tilde{\chi}^0_3\)                         & \(1.044\times 10^{-7}\) & \(7.8\times10^{-2}\)  &  \(1.044\times 10^{-7}\) & \(9.96\times10^{-2}\) &  \(-6.5\times10^{-2}\) \\ 
\(e \bar{e} \rightarrow \tilde{\chi}^0_4 \tilde{\chi}^0_3\)                         & \(2.073\times 10^{-4}\) & \(7.65\times10^{-2}\) &  \(2.075\times 10^{-4}\) & \(9.03\times10^{-2}\) &  \(-1.39\times10^{-1}\) \\
\(e \bar{e} \rightarrow \tilde{\chi}^+_1 \tilde{\chi}^-_1\)                         & \(1.019\times 10^{-3}\) & \(9.53\times10^{-2}\) &  \(1.019\times 10^{-3}\) & \(9.21\times10^{-2}\) &  \(-3.04\times10^{-2}\) \\
\(e \bar{e} \rightarrow \tilde{\chi}^+_2 \tilde{\chi}^-_2\)                         & \(5.817\times 10^{-4}\) & \(7.19\times10^{-2}\) &  \(5.825\times 10^{-4}\) & \(8.44\times10^{-2}\) &  \(-1.34\times10^{-1}\) \\
\(e \bar{e} \rightarrow \tilde{\chi}^+_1 \tilde{\chi}^-_2\)                         & \(8.151\times 10^{-5}\) & \(6.88\times10^{-2}\) &  \(8.148\times 10^{-5}\) & \(7.66\times10^{-2}\) &  \(3.19\times10^{-2}\) \\ 
\(e \bar{e} \rightarrow \tilde{\chi}^-_1 \tilde{\chi}^+_2\)                         & \(8.149\times 10^{-5}\) & \(7.55\times10^{-2}\) &  \(8.149\times 10^{-5}\) & \(8.02\times10^{-2}\) &  \(2.45\times10^{-3}\) \\ 
\(e \bar{e} \rightarrow h Z\)                                                       & \(6.225\times 10^{-5}\) & \(2.86\times10^{-2}\) &  \(6.222\times 10^{-5}\) & \(3.13\times10^{-2}\) &  \(3.53\times10^{-2}\) \\ 
\(e \bar{e} \rightarrow A^0 H\)                                                     & \(4.806\times 10^{-5}\) & \(5.38\times10^{-2}\) &  \(4.809\times 10^{-5}\) & \(5.85\times10^{-2}\) &  \(-6.45\times10^{-2}\) \\
\(e \bar{e} \rightarrow H^- H^+\)                                                   & \(1.503\times 10^{-4}\) & \(8.04\times10^{-2}\) &  \(1.504\times 10^{-4}\) & \(8.37\times10^{-2}\) &  \(-8.05\times10^{-2}\) \\
\(e \bar{e} \rightarrow \tilde{d}_1 \tilde{d}_1^*\)                                 & \(1.836\times 10^{-4}\) & \(6.84\times10^{-2}\) &  \(1.836\times 10^{-4}\) & \(6.48\times10^{-2}\) &  \(4.36\times10^{-3}\) \\ 
\(e \bar{e} \rightarrow \tilde{d}_2 \tilde{d}_2^*\)                                 & \(1.836\times 10^{-4}\) & \(6.84\times10^{-2}\) &  \(1.836\times 10^{-4}\) & \(6.48\times10^{-2}\) &  \(4.36\times10^{-3}\) \\ 
\(e \bar{e} \rightarrow \tilde{d}_3 \tilde{d}_3^*\)                                 & \(1.48\times 10^{-4}\)  & \(6.8\times10^{-2}\)  &  \(1.48\times 10^{-4}\)  & \(6.74\times10^{-2}\) &  \(-3.24\times10^{-2}\) \\
\(e \bar{e} \rightarrow \tilde{d}_4 \tilde{d}_4^*\)                                 & \(4.47\times 10^{-5}\)  & \(7.65\times10^{-2}\) &  \(4.474\times 10^{-5}\) & \(7.97\times10^{-2}\) &  \(-7.67\times10^{-2}\) \\
\(e \bar{e} \rightarrow \tilde{d}_5 \tilde{d}_5^*\)                                 & \(4.47\times 10^{-5}\)  & \(7.65\times10^{-2}\) &  \(4.474\times 10^{-5}\) & \(7.97\times10^{-2}\) &  \(-7.67\times10^{-2}\) \\
\(e \bar{e} \rightarrow \tilde{d}_6 \tilde{d}_6^*\)                                 & \(4.183\times 10^{-5}\) & \(7.21\times10^{-2}\) &  \(4.187\times 10^{-5}\) & \(7.49\times10^{-2}\) &  \(-8.68\times10^{-2}\) \\
\(e \bar{e} \rightarrow \tilde{d}_3 \tilde{d}_6^*\)                                 & \(1.938\times 10^{-5}\) & \(6.65\times10^{-2}\) &  \(1.939\times 10^{-5}\) & \(6.65\times10^{-2}\) &  \(-2.58\times10^{-2}\) \\
\(e \bar{e} \rightarrow \tilde{d}_6 \tilde{d}_3^*\)                                 & \(1.938\times 10^{-5}\) & \(6.65\times10^{-2}\) &  \(1.939\times 10^{-5}\) & \(6.65\times10^{-2}\) &  \(-2.58\times10^{-2}\) \\
\(d \bar{d} \rightarrow \tilde{e}_2 \tilde{e}_2^*\)                                 & \(2.057\times 10^{-5}\) & \(6.77\times10^{-2}\) &  \(2.058\times 10^{-5}\) & \(8.71\times10^{-2}\) &  \(-4.67\times10^{-2}\) \\
\(d \bar{d} \rightarrow \tilde{e}_3 \tilde{e}_3^*\)                                 & \(4.017\times 10^{-6}\) & \(7.12\times10^{-2}\) &  \(4.02\times 10^{-6}\)  & \(7.14\times10^{-2}\) &  \(-7.47\times10^{-2}\) \\
\(d \bar{d} \rightarrow \tilde{e}_4 \tilde{e}_4^*\)                                 & \(5.008\times 10^{-6}\) & \(7.12\times10^{-2}\) &  \(5.014\times 10^{-6}\) & \(7.16\times10^{-2}\) &  \(-1.1\times10^{-1}\) \\ 
\(d \bar{d} \rightarrow \tilde{e}_5 \tilde{e}_5^*\)                                 & \(5.008\times 10^{-6}\) & \(7.12\times10^{-2}\) &  \(5.014\times 10^{-6}\) & \(7.16\times10^{-2}\) &  \(-1.1\times10^{-1}\) \\ 
\(d \bar{d} \rightarrow \tilde{e}_6 \tilde{e}_6\)                                   & \(1.709\times 10^{-5}\) & \(6.79\times10^{-2}\) &  \(1.71\times 10^{-5}\)  & \(6.78\times10^{-2}\) &  \(-7.02\times10^{-2}\) \\
\(d \bar{d} \rightarrow \tilde{e}_3 \tilde{e}_6\)                                   & \(2.235\times 10^{-6}\) & \(6.61\times10^{-2}\) &  \(2.236\times 10^{-6}\) & \(6.67\times10^{-2}\) &  \(-3.58\times10^{-2}\) \\
\(d \bar{d} \rightarrow \tilde{u}_1 \tilde{u}_1^*\)                                 & \(6.384\times 10^{-3}\) & \(7.52\times10^{-2}\) &  \(6.384\times 10^{-3}\) & \(7.52\times10^{-2}\) &  \(0.\) \\ 
\(d \bar{d} \rightarrow \tilde{u}_3 \tilde{u}_3^*\)                                 & \(6.415\times 10^{-3}\) & \(7.51\times10^{-2}\) &  \(6.415\times 10^{-3}\) & \(7.51\times10^{-2}\) &  \(0.\) \\ 
\(d \bar{d} \rightarrow \tilde{u}_4 \tilde{u}_4^*\)                                 & \(6.387\times 10^{-3}\) & \(7.53\times10^{-2}\) &  \(6.387\times 10^{-3}\) & \(7.53\times10^{-2}\) &  \(0.\) \\ 
\(d \bar{d} \rightarrow \tilde{u}_5 \tilde{u}_5^*\)                                 & \(6.387\times 10^{-3}\) & \(7.53\times10^{-2}\) &  \(6.387\times 10^{-3}\) & \(7.53\times10^{-2}\) &  \(0.\) \\ 
\(d \bar{d} \rightarrow \tilde{u}_6 \tilde{u}_6^*\)                                 & \(6.379\times 10^{-3}\) & \(7.5\times10^{-2}\)  &  \(6.379\times 10^{-3}\) & \(7.5\times10^{-2}\)  &  \(0.\) \\ 
\(d \bar{d} \rightarrow \tilde{\chi}^0_1 \tilde{\chi}^0_3\)                         & \(3.654\times 10^{-6}\) & \(8.58\times10^{-2}\) &  \(3.656\times 10^{-6}\) & \(9.35\times10^{-2}\) &  \(-5.71\times10^{-2}\) \\
\(d \bar{d} \rightarrow \tilde{\chi}^0_1 \tilde{\chi}^0_4\)                         & \(3.739\times 10^{-6}\) & \(1.01\times10^{-1}\) &  \(3.739\times 10^{-6}\) & \(1.14\times10^{-1}\) &  \(-6.42\times10^{-3}\) \\
\(d \bar{d} \rightarrow \tilde{\chi}^0_2 \tilde{\chi}^0_2\)                         & \(9.874\times 10^{-5}\) & \(7.02\times10^{-2}\) &  \(9.878\times 10^{-5}\) & \(7.04\times10^{-2}\) &  \(-3.48\times10^{-2}\) \\
\(d \bar{d} \rightarrow \tilde{\chi}^0_4 \tilde{\chi}^0_3\)                         & \(9.752\times 10^{-5}\) & \(7.64\times10^{-2}\) &  \(9.765\times 10^{-5}\) & \(9.03\times10^{-2}\) &  \(-1.34\times10^{-1}\) \\
\(d \bar{d} \rightarrow \tilde{\chi}^0_2 \tilde{\chi}^0_4\)                         & \(3.43\times 10^{-5}\)  & \(8.81\times10^{-2}\) &  \(3.434\times 10^{-5}\) & \(7.31\times10^{-2}\) &  \(-1.42\times10^{-1}\) \\
\(d \bar{d} \rightarrow \tilde{\chi}^+_1 \tilde{\chi}^-_1\)                         & \(3.222\times 10^{-4}\) & \(7.32\times10^{-2}\) &  \(3.226\times 10^{-4}\) & \(7.65\times10^{-2}\) &  \(-9.94\times10^{-2}\) \\
\(d \bar{d} \rightarrow \tilde{\chi}^+_2 \tilde{\chi}^-_2\)                         & \(9.022\times 10^{-5}\) & \(8.71\times10^{-2}\) &  \(9.016\times 10^{-5}\) & \(1.21\times10^{-1}\) &  \(6.44\times10^{-2}\) \\ 
\(d \bar{d} \rightarrow \tilde{\chi}^+_1 \tilde{\chi}^-_2\)                         & \(2.416\times 10^{-5}\) & \(6.46\times10^{-2}\) &  \(2.416\times 10^{-5}\) & \(6.9\times10^{-2}\)  &  \(-2.28\times10^{-2}\) \\ 
\(d \bar{d} \rightarrow \tilde{\chi}^-_1 \tilde{\chi}^+_2\)                         & \(2.415\times 10^{-5}\) & \(7.88\times10^{-2}\) &  \(2.417\times 10^{-5}\) & \(8.02\times10^{-2}\) &  \(-9.11\times10^{-2}\) \\
\(d \bar{d} \rightarrow h Z\)                                                       & \(3.045\times 10^{-5}\) & \(2.86\times10^{-2}\) &  \(3.044\times 10^{-5}\) & \(3.13\times10^{-2}\) &  \(3.61\times10^{-2}\) \\ 
\(d \bar{d} \rightarrow A^0 H\)                                                     & \(2.352\times 10^{-5}\) & \(5.38\times10^{-2}\) &  \(2.353\times 10^{-5}\) & \(5.85\times10^{-2}\) &  \(-6.38\times10^{-2}\) \\
\(d \bar{d} \rightarrow H^- H^+\)                                                   & \(2.049\times 10^{-5}\) & \(6.8\times10^{-2}\)  &  \(2.05\times 10^{-5}\)  & \(8.79\times10^{-2}\) &  \(-3.95\times10^{-2}\) \\
\(d \bar{d} \rightarrow \tilde{d}_1 \tilde{d}_1^*\)                                 & \(9.528\times 10^{-2}\) & \(3.72\times10^{-2}\) &  \(9.528\times 10^{-2}\) & \(3.72\times10^{-2}\) &  \(0.\) \\ 
\(d \bar{d} \rightarrow \tilde{d}_2 \tilde{d}_2^*\)                                 & \(6.383\times 10^{-3}\) & \(7.51\times10^{-2}\) &  \(6.383\times 10^{-3}\) & \(7.51\times10^{-2}\) &  \(0.\) \\ 
\(d \bar{d} \rightarrow \tilde{d}_3 \tilde{d}_3^*\)                                 & \(6.394\times 10^{-3}\) & \(7.53\times10^{-2}\) &  \(6.394\times 10^{-3}\) & \(7.53\times10^{-2}\) &  \(0.\) \\ 
\(d \bar{d} \rightarrow \tilde{d}_4 \tilde{d}_4^*\)                                 & \(9.533\times 10^{-2}\) & \(3.85\times10^{-2}\) &  \(9.533\times 10^{-2}\) & \(3.85\times10^{-2}\) &  \(0.\) \\ 
\(d \bar{d} \rightarrow \tilde{d}_5 \tilde{d}_5^*\)                                 & \(6.388\times 10^{-3}\) & \(7.53\times10^{-2}\) &  \(6.388\times 10^{-3}\) & \(7.53\times10^{-2}\) &  \(0.\) \\ 
\(d \bar{d} \rightarrow \tilde{d}_6 \tilde{d}_6^*\)                                 & \(6.388\times 10^{-3}\) & \(7.53\times10^{-2}\) &  \(6.388\times 10^{-3}\) & \(7.53\times10^{-2}\) &  \(0.\) \\ 
\(\tau \bar{\tau} \rightarrow \tilde{e}_1 \tilde{e}_1^*\)                           & \(1.509\times 10^{-4}\) & \(8.07\times10^{-2}\) &  \(1.51\times 10^{-4}\)  & \(8.88\times10^{-2}\) &  \(-1.11\times10^{-1}\) \\
\(\tau \bar{\tau} \rightarrow \tilde{e}_2 \tilde{e}_2^*\)                           & \(1.509\times 10^{-4}\) & \(8.07\times10^{-2}\) &  \(1.51\times 10^{-4}\)  & \(8.88\times10^{-2}\) &  \(-1.11\times10^{-1}\) \\
\(\tau \bar{\tau} \rightarrow \tilde{e}_3 \tilde{e}_3^*\)                           & \(3.975\times 10^{-3}\) & \(3.25\times10^{-2}\) &  \(3.977\times 10^{-3}\) & \(3.44\times10^{-2}\) &  \(-4.44\times10^{-2}\) \\
\(\tau \bar{\tau} \rightarrow \tilde{e}_4 \tilde{e}_4^*\)                           & \(1.353\times 10^{-4}\) & \(7.8\times10^{-2}\)  &  \(1.354\times 10^{-4}\) & \(8.39\times10^{-2}\) &  \(-1.09\times10^{-1}\) \\
\(\tau \bar{\tau} \rightarrow \tilde{e}_5 \tilde{e}_5^*\)                           & \(1.353\times 10^{-4}\) & \(7.8\times10^{-2}\)  &  \(1.354\times 10^{-4}\) & \(8.39\times10^{-2}\) &  \(-1.09\times10^{-1}\) \\
\(\tau \bar{\tau} \rightarrow \tilde{e}_6 \tilde{e}_6\)                             & \(3.82\times 10^{-3}\)  & \(4.18\times10^{-2}\) &  \(3.821\times 10^{-3}\) & \(3.58\times10^{-2}\) &  \(-3.69\times10^{-2}\) \\
\(\tau \bar{\tau} \rightarrow \tilde{e}_3 \tilde{e}_6\)                             & \(7.905\times 10^{-4}\) & \(4.84\times10^{-2}\) &  \(7.905\times 10^{-4}\) & \(4.84\times10^{-2}\) &  \(3.42\times10^{-9}\) \\
\(\tau \bar{\tau} \rightarrow \tilde{u}_1 \tilde{u}_1^*\)                           & \(2.715\times 10^{-4}\) & \(6.84\times10^{-2}\) &  \(2.716\times 10^{-4}\) & \(8.02\times10^{-2}\) &  \(-5.93\times10^{-2}\) \\
\(\tau \bar{\tau} \rightarrow \tilde{u}_2 \tilde{u}_2^*\)                           & \(2.715\times 10^{-4}\) & \(6.84\times10^{-2}\) &  \(2.716\times 10^{-4}\) & \(8.02\times10^{-2}\) &  \(-5.93\times10^{-2}\) \\
\(\tau \bar{\tau} \rightarrow \tilde{u}_3 \tilde{u}_3^*\)                           & \(1.687\times 10^{-4}\) & \(7.34\times10^{-2}\) &  \(1.689\times 10^{-4}\) & \(7.5\times10^{-2}\)  &  \(-8.3\times10^{-2}\) \\ 
\(\tau \bar{\tau} \rightarrow \tilde{u}_4 \tilde{u}_4^*\)                           & \(1.788\times 10^{-4}\) & \(7.65\times10^{-2}\) &  \(1.79\times 10^{-4}\)  & \(8.57\times10^{-2}\) &  \(-9.84\times10^{-2}\) \\ 
\(\tau \bar{\tau} \rightarrow \tilde{u}_5 \tilde{u}_5^*\)                           & \(1.788\times 10^{-4}\) & \(7.65\times10^{-2}\) &  \(1.79\times 10^{-4}\)  & \(8.57\times10^{-2}\) &  \(-9.84\times10^{-2}\) \\ 
\(\tau \bar{\tau} \rightarrow \tilde{u}_6 \tilde{u}_6^*\)                           & \(2.036\times 10^{-4}\) & \(7.1\times10^{-2}\)  &  \(2.038\times 10^{-4}\) & \(7.79\times10^{-2}\) &  \(-1.1\times10^{-1}\) \\ 
\(\tau \bar{\tau} \rightarrow \tilde{u}_3 \tilde{u}_6^*\)                           & \(3.949\times 10^{-5}\) & \(6.65\times10^{-2}\) &  \(3.949\times 10^{-5}\) & \(6.65\times10^{-2}\) &  \(-7.06\times10^{-4}\) \\
\(\tau \bar{\tau} \rightarrow \tilde{u}_6 \tilde{u}_3^*\)                           & \(3.949\times 10^{-5}\) & \(6.65\times10^{-2}\) &  \(3.949\times 10^{-5}\) & \(6.65\times10^{-2}\) &  \(-7.06\times10^{-4}\) \\
\(\tau \bar{\tau} \rightarrow \tilde{\chi}^0_1 \tilde{\chi}^0_1\)                   & \(6.338\times 10^{-4}\) & \(7.74\times10^{-2}\) &  \(6.334\times 10^{-4}\) & \(7.32\times10^{-2}\) &  \(6.12\times10^{-2}\) \\ 
\(\tau \bar{\tau} \rightarrow \tilde{\chi}^0_1 \tilde{\chi}^0_3\)                   & \(1.229\times 10^{-5}\) & \(1.23\times10^{-1}\) &  \(1.231\times 10^{-5}\) & \(1.27\times10^{-1}\) &  \(-1.63\times10^{-1}\) \\
\(\tau \bar{\tau} \rightarrow \tilde{\chi}^0_1 \tilde{\chi}^0_4\)                   & \(3.056\times 10^{-5}\) & \(1.2\times10^{-1}\)  &  \(3.063\times 10^{-5}\) & \(1.2\times10^{-1}\)  &  \(-2.23\times10^{-1}\) \\ 
\(\tau \bar{\tau} \rightarrow \tilde{\chi}^0_2 \tilde{\chi}^0_2\)                   & \(4.228\times 10^{-4}\) & \(8.23\times10^{-2}\) &  \(4.233\times 10^{-4}\) & \(8.13\times10^{-2}\) &  \(-1.21\times10^{-1}\) \\
\(\tau \bar{\tau} \rightarrow \tilde{\chi}^0_3 \tilde{\chi}^0_3\)                   & \(1.007\times 10^{-6}\) & \(7.48\times10^{-2}\) &  \(1.007\times 10^{-6}\) & \(7.97\times10^{-2}\) &  \(-6.75\times10^{-2}\) \\
\(\tau \bar{\tau} \rightarrow \tilde{\chi}^0_4 \tilde{\chi}^0_3\)                   & \(1.854\times 10^{-4}\) & \(7.57\times10^{-2}\) &  \(1.853\times 10^{-4}\) & \(7.56\times10^{-2}\) &  \(3.07\times10^{-2}\) \\ 
\(\tau \bar{\tau} \rightarrow \tilde{\chi}^+_1 \tilde{\chi}^-_1\)                   & \(1.019\times 10^{-3}\) & \(1.\times10^{-1}\)   &  \(1.018\times 10^{-3}\) & \(7.47\times10^{-2}\) &  \(1.18\times10^{-1}\) \\ 
\(\tau \bar{\tau} \rightarrow \tilde{\chi}^+_2 \tilde{\chi}^-_2\)                   & \(5.731\times 10^{-4}\) & \(7.08\times10^{-2}\) &  \(5.736\times 10^{-4}\) & \(8.54\times10^{-2}\) &  \(-8.84\times10^{-2}\) \\
\(\tau \bar{\tau} \rightarrow \tilde{\chi}^+_1 \tilde{\chi}^-_2\)                   & \(8.026\times 10^{-5}\) & \(9.44\times10^{-2}\) &  \(8.028\times 10^{-5}\) & \(7.75\times10^{-2}\) &  \(-3.6\times10^{-2}\) \\ 
\(\tau \bar{\tau} \rightarrow \tilde{\chi}^-_1 \tilde{\chi}^+_2\)                   & \(8.02\times 10^{-5}\)  & \(1.12\times10^{-1}\) &  \(8.027\times 10^{-5}\) & \(9.7\times10^{-2}\)  &  \(-8.74\times10^{-2}\) \\ 
\(\tau \bar{\tau} \rightarrow h h\)                                                 & \(1.362\times 10^{-9}\) & \(1.99\times10^{-1}\) &  \(1.364\times 10^{-9}\) & \(1.75\times10^{-1}\) &  \(-1.69\times10^{-1}\) \\
\(\tau \bar{\tau} \rightarrow H H\)                                                 & \(7.443\times 10^{-7}\) & \(1.71\times10^{-1}\) &  \(7.434\times 10^{-7}\) & \(1.58\times10^{-1}\) &  \(1.22\times10^{-1}\) \\ 
\(\tau \bar{\tau} \rightarrow h H\)                                                 & \(1.085\times 10^{-7}\) & \(4.58\times10^{-2}\) &  \(1.084\times 10^{-7}\) & \(4.32\times10^{-2}\) &  \(3.69\times10^{-2}\) \\ 
\(\tau \bar{\tau} \rightarrow h Z\)                                                 & \(6.48\times 10^{-5}\)  & \(1.31\times10^{-1}\) &  \(6.48\times 10^{-5}\)  & \(1.31\times10^{-1}\) &  \(1.54\times 10^{-5}\) \\ 
\(\tau \bar{\tau} \rightarrow h \gamma\)                                            & \(2.549\times 10^{-6}\) & \(2.39\times10^{-2}\) &  \(2.549\times 10^{-6}\) & \(2.39\times10^{-2}\) &  \(0.\) \\ 
\(\tau \bar{\tau} \rightarrow A^0 H\)                                               & \(5.287\times 10^{-5}\) & \(5.54\times10^{-2}\) &  \(5.286\times 10^{-5}\) & \(5.72\times10^{-2}\) &  \(9.84\times10^{-3}\) \\ 
\(\tau \bar{\tau} \rightarrow A^0 Z\)                                               & \(2.663\times 10^{-3}\) & \(1.61\times10^{-1}\) &  \(2.663\times 10^{-3}\) & \(1.61\times10^{-1}\) &  \(3.76\times10^{-9}\) \\
\(\tau \bar{\tau} \rightarrow A^0 \gamma\)                                          & \(1.957\times 10^{-4}\) & \(2.43\times10^{-2}\) &  \(1.957\times 10^{-4}\) & \(2.43\times10^{-2}\) &  \(0.\) \\ 
\(\tau \bar{\tau} \rightarrow H^- W^+\)                                             & \(3.124\times 10^{-3}\) & \(1.14\times10^{-1}\) &  \(3.123\times 10^{-3}\) & \(1.24\times10^{-1}\) &  \(1.92\times10^{-2}\) \\ 
\(\tau \bar{\tau} \rightarrow H^- H^+\)                                             & \(1.497\times 10^{-4}\) & \(8.44\times10^{-2}\) &  \(1.499\times 10^{-4}\) & \(9.17\times10^{-2}\) &  \(-1.04\times10^{-1}\) \\
\(\tau \bar{\tau} \rightarrow \tilde{d}_1 \tilde{d}_1^*\)                           & \(1.836\times 10^{-4}\) & \(6.84\times10^{-2}\) &  \(1.836\times 10^{-4}\) & \(7.03\times10^{-2}\) &  \(-1.09\times10^{-3}\) \\
\(\tau \bar{\tau} \rightarrow \tilde{d}_2 \tilde{d}_2^*\)                           & \(1.836\times 10^{-4}\) & \(6.84\times10^{-2}\) &  \(1.836\times 10^{-4}\) & \(7.03\times10^{-2}\) &  \(-1.09\times10^{-3}\) \\
\(\tau \bar{\tau} \rightarrow \tilde{d}_3 \tilde{d}_3^*\)                           & \(1.48\times 10^{-4}\)  & \(6.8\times10^{-2}\)  &  \(1.48\times 10^{-4}\)  & \(6.95\times10^{-2}\) &  \(-3.24\times10^{-2}\) \\
\(\tau \bar{\tau} \rightarrow \tilde{d}_4 \tilde{d}_4^*\)                           & \(4.47\times 10^{-5}\)  & \(7.65\times10^{-2}\) &  \(4.475\times 10^{-5}\) & \(8.57\times10^{-2}\) &  \(-1.01\times10^{-1}\) \\
\(\tau \bar{\tau} \rightarrow \tilde{d}_5 \tilde{d}_5^*\)                           & \(4.47\times 10^{-5}\)  & \(7.65\times10^{-2}\) &  \(4.475\times 10^{-5}\) & \(8.57\times10^{-2}\) &  \(-1.01\times10^{-1}\) \\
\(\tau \bar{\tau} \rightarrow \tilde{d}_6 \tilde{d}_6^*\)                           & \(4.184\times 10^{-5}\) & \(7.21\times10^{-2}\) &  \(4.187\times 10^{-5}\) & \(7.5\times10^{-2}\)  &  \(-8.46\times10^{-2}\) \\ 
\(\tau \bar{\tau} \rightarrow \tilde{d}_3 \tilde{d}_6^*\)                           & \(1.94\times 10^{-5}\)  & \(6.65\times10^{-2}\) &  \(1.94\times 10^{-5}\)  & \(6.65\times10^{-2}\) &  \(-2.08\times10^{-4}\) \\
\(\tau \bar{\tau} \rightarrow \tilde{d}_6 \tilde{d}_3^*\)                           & \(1.94\times 10^{-5}\)  & \(6.65\times10^{-2}\) &  \(1.94\times 10^{-5}\)  & \(6.65\times10^{-2}\) &  \(-2.08\times10^{-4}\) \\
\(d \bar{d} \rightarrow \tilde{g} \tilde{g}\)                                       & \(6.229\times 10^{-2}\) & \(7.46\times10^{-2}\) &  \(6.23\times 10^{-2}\)  & \(7.4\times10^{-2}\)  &  \(-3.21\times10^{-3}\) \\ 
\(d \bar{d} \rightarrow g g\)                                                       & \(8.497\times 10^{-1}\) & \(2.67\times10^{-2}\) &  \(8.497\times 10^{-1}\) & \(2.67\times10^{-2}\) &  \(0.\) \\ 
\(u \bar{u} \rightarrow \tilde{g} \tilde{g}\)                                       & \(6.23\times 10^{-2}\)  & \(7.46\times10^{-2}\) &  \(6.23\times 10^{-2}\)  & \(7.4\times10^{-2}\)  &  \(-1.61\times10^{-3}\) \\ 
\(g g \rightarrow \tilde{g} \tilde{g}\)                                             & \(9.734\times 10^{-1}\) & \(2.97\times10^{-2}\) &  \(9.734\times 10^{-1}\) & \(2.97\times10^{-2}\) &  \(0.\) \\ 
\(\bar{d} \tilde{d}_1 \rightarrow \tilde{g} g\)                                     & \(9.162\times 10^{-1}\) & \(3.86\times10^{-2}\) &  \(9.162\times 10^{-1}\) & \(3.86\times10^{-2}\) &  \(0.\) \\ 
\(g g \rightarrow \tilde{d}_2 \tilde{d}_2^*\)                                       & \(8.585\times 10^{-3}\) & \(9.07\times10^{-2}\) &  \(8.585\times 10^{-3}\) & \(9.07\times10^{-2}\) &  \(0.\) \\ 
\(g g \rightarrow \tilde{u}_3 \tilde{u}_3^*\)                                       & \(8.788\times 10^{-3}\) & \(9.17\times10^{-2}\) &  \(8.788\times 10^{-3}\) & \(9.17\times10^{-2}\) &  \(0.\) \\ 
\(g g \rightarrow b \bar{b}\)                                                       & \(4.104\times 10^{-1}\) & \(2.29\times10^{-2}\) &  \(4.104\times 10^{-1}\) & \(2.29\times10^{-2}\) &  \(0.\) \\ 
\(g g \rightarrow t \bar{t}\)                                                       & \(2.033\times 10^{-1}\) & \(2.7\times10^{-2}\)  &  \(2.033\times 10^{-1}\) & \(2.7\times10^{-2}\)  &  \(0.\) \\ 
\hline 
\end{longtable}

\subsection{Parameter and run card}
\label{app:cards}
We used for the calculation of the cross sections the run and parameter card delivered with {\tt MadGraph 5.1.4.8.4}. However, in order to reduce the numerical uncertainity, we changed the following entries in the run card \footnote{The interested reader who needs the full run and parameter card in order to reproduce this comparison, can contact the author via email}. 
\begin{verbatim}
...
#*********************************************************************
# Collider type and energy                                           *
#*********************************************************************
         0     = lpp1  ! beam 1 type (0=NO PDF)
         0     = lpp2  ! beam 2 type (0=NO PDF)
...
#*********************************************************************
# Renormalization and factorization scales                           *
#*********************************************************************
 T        = fixed_ren_scale  ! if .true. use fixed ren scale
 T        = fixed_fac_scale  ! if .true. use fixed fac scale
...
\end{verbatim}

\end{appendix}

\bibliographystyle{ieeetr}

\end{document}